\documentclass[fleqn,usenatbib]{mnras}
\usepackage{newtxtext,newtxmath}
\usepackage[T1]{fontenc}

\DeclareRobustCommand{\VAN}[3]{#2}
\let\VANthebibliography\thebibliography
\def\thebibliography{\DeclareRobustCommand{\VAN}[3]{##3}\VANthebibliography}

\usepackage{graphicx}	
\usepackage{amsmath}	
\usepackage{booktabs}   
\usepackage{caption}
\usepackage{subcaption}
\usepackage[flushleft]{threeparttable}
\usepackage{comment}

\DeclareGraphicsExtensions{.jpg,.png}

\title[AGN impact on CO emission]{
AGN impact on the molecular gas in galactic centers
as probed by CO lines}

\author[F. Esposito et al.]{
Federico Esposito$^{1,2}$\thanks{E-mail: federico.esposito7@unibo.it},
Livia Vallini$^{3}$,
Francesca Pozzi$^{1,2}$,
Viviana Casasola$^{4}$,
Matilde Mingozzi$^{5}$,
\newauthor{
Cristian Vignali$^{1,2}$,
Carlotta Gruppioni$^{2}$
and Francesco Salvestrini$^{6}$}
\\
$^{1}$Dipartimento di Fisica e Astronomia, Università degli Studi di Bologna, Via P. Gobetti 93/2, I-40129 Bologna, Italy\\
$^{2}$Osservatorio di Astrofisica e Scienza dello Spazio (INAF–OAS), Via P. Gobetti 93/3, I-40129 Bologna, Italy\\
$^{3}$Scuola Normale Superiore, Piazza dei Cavalieri 7, I-56126 Pisa, Italy\\
$^{4}$INAF – Istituto di Radioastronomia, Via P. Gobetti 101, 40129, Bologna, Italy\\
$^{5}$Space Telescope Science Institute, 3700 San Martin Drive, Baltimore, MD 21218, USA \\
$^{6}$INAF - Osservatorio Astrofisico di Arcetri, Largo Enrico Fermi 5, 50125, Firenze, Italy \\
}

\date{Accepted XXX. Received YYY; in original form ZZZ}

\pubyear{2021}

\begin{document}
\label{firstpage}
\pagerange{\pageref{firstpage}--\pageref{lastpage}}
\maketitle

\begin{abstract}
We present a detailed analysis
of the X-ray, infrared, and
carbon monoxide (CO) emission
for a sample of 35 local 
($z \leq 0.15$), active
($L_{\text{X}} \geq 10^{42}$ erg s$^{-1}$)
galaxies.
Our goal is to infer the contribution of
far-ultraviolet (FUV) radiation from star formation (SF),
and X-ray radiation from the active galactic nuclei (AGN),
respectively producing
photodissociation regions (PDRs)
and X-ray dominated regions (XDRs), 
to the molecular gas heating. 
To this aim, we exploit the CO spectral line energy
distribution (CO SLED) as traced by Herschel, 
complemented with data from single-dish telescopes 
for the low-$J$ lines, 
and high-resolution ALMA images of the mid-$J$ 
CO emitting region. 
By comparing our results to the
Schmidt-Kennicutt relation, 
we find no evidence for AGN influence
on the cold and low-density gas on kpc-scales. 
On nuclear ($r = 250$ pc) scales,
we find weak correlations
between the CO line ratios
and either the FUV
or X-ray fluxes:
this may indicate that neither SF nor AGN radiation
dominates the gas excitation,
at least at $r = 250$ pc.
From a comparison of the CO line ratios
with PDR and XDR models,
we find that PDRs can reproduce
observations only in presence of extremely high gas densities
($n > 10^5$ cm$^{-3}$). In the XDR case, instead, the models
suggest moderate densities
($n \approx 10^{2-4}$ cm$^{-3}$).
We conclude that a mix of the two mechanisms
(PDR for the mid-$J$, XDR 
or possibly shocks 
for the high-$J$)
is necessary to explain the 
observed CO excitation in active galaxies.
\end{abstract}


\section{Introduction}

The molecular gas phase of the 
interstellar medium (ISM) is 
the fuel for star formation (SF), thus
it plays a central role in galaxy evolution
\citep{mckee07,carilli13,tacconi20}.
At the same time, the molecular gas properties
(e.g. temperature, density, turbulence,
chemical composition) are affected 
by feedback processes induced by SF and
by the accretion onto the central black hole
in sources hosting an active galactic nucleus (AGN)
\citep{aalto95, omont07, imanishi11, imanishi16}.
A key question is whether, and on which spatial scales, 
the effect of AGN radiation on the molecular gas 
can produce observable effects 
that can be retrieved from the molecular line emission.

Molecular hydrogen (H$_2$), dominating the mass 
of molecular ISM, does not have a dipole moment, 
and the quadrupole transitions require 
high temperatures ($T=500-2000\rm \, K$), mainly
present in shock-heated gas
\citep{flower10}. 
For this reason, the most used molecular gas tracer is the
carbon monoxide (CO) which
has instead bright dipole emission and is 
the second most abundant molecule in the Universe 
\citep{bolatto13}.

Moreover, the CO Spectral Line Energy Distribution (CO SLED),
i.e. the luminosity of CO rotational lines as a function 
of the rotational quantum number 
$J$\footnote{The CO SLED is also often referred to 
as the \textit{CO rotational ladder}.},
is a very powerful diagnostic for
the physical conditions of molecular ISM
\citep{narayanan14, rosenberg15}.
The CO SLED can be broken down into three 
different parts \citep[e.g.][]{vallini19,decarli20}.
The low-$J$ lines ($J_{\text{upp}} \leq 3$)
trace the cold ($T \approx 20 - 50$ K),
low-density ($n \lesssim 10^3$ cm$^{-3}$)
gas; this is where the majority
of the mass resides, 
so these lines are good tracers of the total
molecular gas mass in galaxies \citep[][]{bolatto13}.
Both the mid-$J$ ($4 \leq J_{\text{upp}} \leq 7$)
and the high-$J$ ($ J_{\text{upp}} \geq 8$) lines
originate from increasingly denser
($n \approx 10^4 - 10^6$ cm$^{-3}$) 
and warmer ($T \approx 100 - 500$ K)
molecular gas \citep{greve14}. 
For this reason, the excitation of the CO ladder, 
especially in the mid/high-$J$ part, 
can be exploited to disentangle different heating sources
such as radiation from SF, AGN accretion, 
and mechanical heating from shocks
\citep[e.g.][]{vanderwerf10, mingozzi18}.

Stellar radiation affects the molecular gas
mainly in the far-ultraviolet
(FUV, $6<h\nu<13.6\rm \, eV$) band,
where photons can dissociate H$_2$ molecules
without ionizing H atoms
(for which photons with $h \nu \geq 13.6$ eV are needed). 
The FUV photon penetration creates a transition layer,
called photodissociation region (PDR),
linking the outer 
HII region and the fully molecular layers
of Giant Molecular Clouds (GMCs).
FUV-induced photoelectric effect on dust grains 
is the major heating mechanism in PDRs
\citep{hollenbach97, hollenbach99}, 
which then cool down through metal fine structure line 
emission (e.g. [CII] $158\,\rm \mu m$, [OI] $63\,\rm \mu m$) 
and molecular rotational lines, among which CO transitions.
The FUV flux is usually parametrized 
in terms of the Habing field
($G_0 = 1.6 \times 10^{-3}$ erg s$^{-1}$ cm$^{-2}$,
\citealt{habing68}).

X-ray photons from the AGN penetrate deeper than FUV photons
into the molecular clouds and 
create the so-called X-ray Dominated Regions
\citep[XDR;][]{maloney96}. 
There, the heating and chemical composition of the gas 
are peculiarly influenced by the 
$\sim$1--100 keV X-ray radiation
\citep{maloney96, meijerink05, meijerink07},
keeping the molecular gas warmer 
at larger (column) densities,
following the release of fast photoelectrons
\citep{morrison83, wilms00}.

PDR and XDR models are radiative transfer calculations
\citep{hollenbach99, meijerink07, ferland17}
that take the impinging radiation 
(FUV and X-ray photons, respectively), the gas density, column density, and metallicity as input, 
and return the expected line emission.
While the low/mid-$J$ CO emission is usually consistent
with the presence of a PDR component produced by SF 
\citep{pereira13, kame14, talia18}, 
in active galaxies with peculiarly excited high-$J$ CO lines
\citep{vanderwerf10, schleicher10, gallerani14, pozzi17, vallini19, pensabene21} 
an XDR, associated with the AGN activity, 
is often necessary to reproduce the CO SLEDs. 

The purpose of this work is to investigate
the possible relation between the AGN activity 
and the conditions of molecular gas in a sample of local active galaxies with well-sampled CO SLED.
We will assess whether, and to what extent, 
the excitation of the CO ladder 
shows correlations with X-ray and FUV tracers 
and whether the CO SLED can be used to infer 
the effect of SF versus AGN heating 
on the whole host galaxy 
and within the nuclear region.

The paper is structured as follows:
in Section~\ref{sec:sample}
we introduce the sample
and the selection criteria.
In Section~\ref{sec:data}
we describe the data collection
from the sub-mm up to the X-ray band.
In Section~\ref{sec:SK}
we derive the CO emission on a galactic scale,
and we study the Schmidt--Kennicutt relation.
In Section~\ref{sec:results} we derive
the physical parameters for the PDR and XDR
analysis and we discuss the results we find.
We assume a $\Lambda$CDM cosmology
with $H_0 = 70$ km s$^{-1}$ Mpc$^{-1}$,
$\Omega_m = 0.3$ and $\Omega_{\Lambda} = 0.7$.

\section{Sample selection}
\label{sec:sample}

\begin{table*}
\centering
\caption{Properties of the sample of 35 AGN.}
\label{tab:sample}
\begin{threeparttable}
\begin{tabular}{llllllllllll}
\toprule
{} & RA & Dec & $D_{\rm L}$ & D$_{25}$ & Class & log$L_{\rm X}$ & log$N_{\rm H}$ & logL$_{\text{IR}}$ & logM$_{\text{mol}}$ & SFR & Sample \\
Name & (deg) & (deg) & (Mpc) & ($''$) & & (erg s$^{-1}$) & (cm$^{-2}$) & (L$_{\odot}$) & (M$_{\odot}$) & (M$_{\odot}$ yr$^{-1}$) &  \\
\midrule
NGC 0034                      &    2.78 &  -12.11 &   85              &    69              & S2   & 42.11 &  23.72 &  11.44              &   9.97              &   31 &     klr-sn \\
UGC 00545                     &   13.40 &   12.69 &  264              &    29              & Q    & 43.60 &     -- &  11.95$^{\text{d}}$ &  10.17              &   34 &        k-n \\
NGC 1068                      &   40.67 &   -0.01 &   16 &   370              & S1h  & 42.38 &  24.70 &  11.27              &  10.14              &   17 &   klmr-cbn \\
3C 84                         &   49.95 &   41.51 &   76              &   128              & S1.5 & 43.98 &  21.68 &  11.20              &   9.63              &  9.0 &       kl-b \\
NGC 1365                      &   53.40 &  -36.14 &   23              &   721              & S1.8 & 42.32 &  22.21 &  11.00              &  10.10              &   17 &      kr-bn \\
IRAS F05189-2524              &   80.26 &  -25.36 &  188              &    30              & S1h  & 43.20 &  22.86 &  12.11              &  10.04              &  109 &   klpr-sbn \\
IRAS 07598+6508               &  121.14 &   65.00 &  704              &    39$^{\text{c}}$ & S1   & 42.10 &     -- &  12.46$^{\text{e}}$ &  10.54              &   -- &       kp-n \\
UGC 05101                     &  143.97 &   61.35 &  174              &    72              & S1   & 43.08 &  24.08 &  11.95              &  10.21              &  105 &    klp-xbn \\
NGC 3227                      &  155.88 &   19.87 &   17 &   239              & S    & 42.10 &  20.95 &  10.13              &   9.02              & 0.56 &       k-bn \\
NGC 4151                      &  182.64 &   39.41 &  14 &   173              & S    & 42.31 &  22.71 &  10.20              &   7.42              & 0.25 &       k-bn \\
NGC 4388                      &  186.45 &   12.66 &   36              &   322              & S1h  & 42.60 &  23.50 &  10.00              &   9.40              &  3.7 &      k-sbn \\
TOL 1238-364                  &  190.22 &  -36.76 &   47              &    76              & S1h  & 43.40 &  24.95 &  10.62              &   8.94              &  4.1 &        k-s \\
Mrk 0231                      &  194.06 &   56.87 &  186              &    85              & S1   & 42.50 &  22.85 &  12.51              &  10.39              &  278 &    klpmr-n \\
MCG -03-34-064                &  200.60 &  -16.73 &   72              &    81              & S1h  & 43.18 &  23.80 &  11.24              &     --              &  5.7 &     kl-sbn \\
NGC 5128                      &  201.37 &  -43.02 &  8 &  1542              & S2$^{\text{i}}$ & 42.39 &  23.02 &  10.11              &  10.17              &  6.7 &        k-b \\
NGC 5135                      &  201.43 &  -29.83 &   59              &   144              & S2   & 41.97 &  24.47 &  11.17              &  10.17              &   17 &      rlk-s \\
Mrk 0463                      &  209.01 &   18.37 &  224              &    64              & S1h  & 43.28 &  23.83 &  11.77$^{\text{e}}$ &   9.92              &   -- &     kp-sbn \\
IC 4518a                      &  224.42 &  -43.13 &   71              &    55              & S2   & 42.64 &  23.36 &  11.13              &     --              &  5.6 &       kl-b \\
VV 705$^{\text{a}}$           &  229.53 &   42.75 &  177              &    39              & S2$^{\text{i}}$ & 42.30 &  23.93 &  11.89              &  10.37              &   72 &       kl-n \\
PKS 1549-79                   &  239.25 &  -79.24 &  725              &    --              & S1i  & 44.71 &  20.00 &  12.36$^{\text{d}}$ &  10.01$^{\text{h}}$ &   -- &        k-b \\
PG 1613+658                   &  243.49 &   65.72 &  605              &    27              & Q    & 44.19 &  20.00 &  12.00              &  10.24              &   44 &        k-b \\
NGC 6240                      &  253.25 &    2.40 &  107              &   131              & S3   & 43.58 &  24.20 &  11.85              &  10.58              &   70 &  klpmr-cbn \\
IRAS 19254-7245$^{\text{b}}$  &  292.84 &  -72.66 &  277              &    38              & S2$^{\text{i}}$ & 42.80 &  23.58 &  12.06$^{\text{e}}$ &  10.34              &  104 &       kp-n \\
3C 405                        &  299.87 &   40.73 &  250              &    33              & S1.9 & 44.37 &  23.38 &  <11.75$^{\text{g}}$ &   <8.88              &   35 &        k-b \\
MCG +04-48-002                &  307.15 &   25.73 &   60              &    60              & S2$^{\text{i}}$ & 43.13 &  23.86 &  11.06              &   9.64              &   10 &       kl-b \\
IC 5063                       &  313.01 &  -57.07 &   49              &   161              & S1h  & 42.87 &  23.42 &  10.85              &   9.36              &  2.6 &       k-sb \\
ESO 286-IG019                 &  314.61 &  -42.65 &  190              &    41              & H2   & 42.30 &  23.69 &  12.00              &  10.25              &  105 &      klp-n \\
3C 433                        &  320.94 &   25.07 &  468              &    19              & S2   & 44.16 &  23.01 &  <11.66$^{\text{g}}$ &   <9.71              &   10 &        k-b \\
NGC 7130                      &  327.08 &  -34.95 &   70              &    93              & S1.9 & 42.30 &  24.10 &  11.35              &  10.10              &   22 &     kl-scb \\
NGC 7172                      &  330.51 &  -31.87 &   37              &   151              & S2   & 42.76 &  22.91 &  10.45              &   9.58              &  2.5 &       k-bn \\
NGC 7465                      &  345.50 &   15.97 &   28              &   64               & S3   & 41.97 &  21.46 &  10.10              &   8.88              & 0.76 &        k-b \\
NGC 7469                      &  345.82 &    8.87 &   71              &    83              & S    & 43.19 &  20.53 &  11.59              &  10.09              &   35 &     klr-bn \\
ESO 148-IG002                 &  348.95 &  -59.05 &  198              &    56              & H2   & 43.20 &     -- &  12.00              &  10.05              &  108 &      klp-n \\
NGC 7582                      &  349.60 &  -42.37 &   23              &   415              & S1i  & 42.53 &  24.20 &  10.87              &   9.64              &  7.1 &      k-cbn \\
NGC 7674                      &  351.99 &    8.78 &  127              &    67              & S1h  & 43.60 &     -- &  11.50              &  10.46              &   15 &       kl-n \\
\bottomrule
\end{tabular}
\begin{tablenotes}
\item \textbf{Notes.}
RA, Dec from NED.
$D_{\rm L}$ is the luminosity distance, calculated from
the redshift (taken from NED)
according to the adopted cosmology.
D$_{25}$ is the optical diameter, measured
at the isophotal level 25 mag arcsec$^{-2}$ in the B-band,
taken from HyperLEDA.
Class is the AGN 
classification from HyperLEDA:
Q = quasar, 
S1 = broad-line Seyfert 1,
S1i = S1 with a broad Paschen H$\beta$ line,
S1h = S2 which show S1 like spectra in polarized light,
S2 = Seyfert 2,
S1.5 = Seyfert 1.5,
S1.8 = Seyfert 1.8,
S1.9 = Seyfert 1.9,
S = AGN objects without classification,
S3 = LINERs,
H2 = extragalactic HII regions.
$L_{\rm X}$ is the 2--10 keV
intrinsic (i.e. corrected for source absorption) luminosity,
taken from the works indicated in the Sample column
(see Section~\ref{sec:xsample} for details).
L$_{\text{IR}}$ is the 8--1000 $\mu$m luminosity,
from \cite{sanders03} unless otherwise specified.
M$_{\text{mol}}$ is the total molecular mass,
calculated as described in Section~\ref{sec:lowj}.
SFR is the star formation rate,
calculated as described in Section~\ref{sec:pdr}.
Sample lists the references for the CO \textit{Herschel} fluxes
($r$ for \cite{rosenberg15}, $m$ for \cite{mashian15},
$p$ for \cite{pearson16}, $k$ for \cite{kame16}, $l$ for \cite{lu17})
and for the X-ray data
($n$ for \cite{brightman11}, $b$ for \cite{ricci17},
$c$ for \cite{marchesi19}, $x$ for \cite{lacaria19},
$s$ for Salvestrini et al. (2022, in prep.)).

\textbf{Additional notes.}
(\textit{a}) RA, Dec from \cite{kojoian81}.
(\textit{b}) RA, Dec from \cite{westmoquette12}.
(\textit{c}) D$_{25}$ from NED.
(\textit{d}) L$_{\text{IR}}$ from \cite{moshir90}.
(\textit{e}) L$_{\text{IR}}$ from \cite{pearson16}.
(\textit{f}) L$_{\text{IR}}$ from the IRAS PSC (1988).
(\textit{g}) Upper limit for L$_{\text{IR}}$ from \cite{golombek88}.
(\textit{h}) M$_{\text{H}_2}$ from \cite{oosterloo19}.
(\textit{i}) Class from NED.

\end{tablenotes}
\end{threeparttable}
\end{table*}

To investigate the impact of AGN activity
onto the molecular gas, we select 
a sample of local galaxies 
adopting the following criteria: 
(\textit{i}) a properly sampled CO SLED 
in the mid/high-$J$ regimes
from \textit{Herschel} observations;
(\textit{ii}) an intrinsic $2-10$ keV luminosity 
L$_X \geq 10^{42}$ erg s$^{-1}$.
Moreover, we collect
low/mid-$J$ CO data 
by considering both sub-mm/mm single-dish observations, 
and interferometric ALMA data, 
which ensure a high spatial resolution.

Selecting sources with intrinsic 
L$_{\text{X}} \geq 10^{42}$ erg s$^{-1}$
is the standard criterion for identifying AGN,
since stellar processes alone
(e.g. X-ray binaries, hot ionized ISM)
rarely reach this X-ray luminosity
\citep[][]{hickox18}.
We look for AGN with a well-sampled CO SLED,
to be able to study the high-$J$ lines
($J_{\text{upp}} \geq 8$),
where we expect to find the imprint of the 
AGN influence on the molecular gas.

The adopted criteria lead to a sample of 
35 active galaxies 
(see Table~\ref{tab:sample}),
with redshifts in the range $0.0015<z<0.15$
(median $z = 0.02$),
corresponding to luminosity distances ($D_{\rm L}$)
in the range $4-720$ Mpc.

Considering the classification 
from the optical spectra, 
$92 \%$ of our AGN 
are classified as Seyfert galaxies and 
two (VV 705 and ESO186--IG019) as 
low-ionisation nuclear emission line regions
(LINERs).
One of our sources (PKS 1549--79) is a quasar
(see \citealt{netzer15} 
for a review on AGN classification), while
PKS 1549-79, 3C84 (Perseus A, NGC 1275), 
3C405 (Cygnus A), and 3C433
are also known as radio sources.

The $8-1000$ $\mu$m infrared luminosities $L_{\text{IR}}$
\citep[from][]{sanders03} cover the range
$10^{10} L_{\odot} < L_{\text{IR}} <
10^{12.5} L_{\odot}$.
The bulk ($43 \%$) of our sample is made of
luminous infrared galaxies (LIRGs, 
$10^{11} \leq L_{\text{IR}} / L_{\odot} < 10^{12} $),
while ultra-luminous infrared
galaxies (ULIRGs,
$L_{\text{IR}} \geq 10^{12} L_{\odot}$) 
account for $27 \%$ of the sample;
the remaining $30 \%$ have 
$10^{10} < L_{\text{IR}} < 10^{11} L_{\odot}$.
It is thought that the (U)LIRG phenomenon
is mainly linked to merger activity
\citep{lonsdale06}, 
especially for 
$L_{\text{IR}} \geq 10^{11.5}$ L$_{\odot}$
\citep{hung14, perez21},
as during mergers
the gas can reach very high gas densities,
triggering intense SF \citep{larson78}.
Mergers and interactions
can also trigger AGN activity for the very same reason:
the gas has the opportunity to lose
its angular momentum and fall from kpc-scale
distances to the inner parsecs from the nucleus
\citep{alonsoherrero12, treister12, ricci17b, ellison19}.
Both SF and AGN phenomena heat the dust,
hence boosting the IR luminosity of the host galaxies.
Within our sample, at least five galaxies
show an evolved merging phase:
ESO~148-IG002 \citep{leslie14},
IRAS~19254-7245 \citep[Superantennae,][]{bendo09},
NGC~6240 \citep{komossa03},
Mrk~463 \citep{bianchi08}
and VV~705 \citep{perna19}.
Seven more galaxies have a very close companion:
NGC~3227 \citep[$\sim$15 kpc,][]{mundell04},
NGC~7465 \citep[$\sim$15 kpc,][]{merkulova12},
NGC~7469 \citep[$\sim$20 kpc,][]{zaragozacardiel17},
NGC~7674 \citep[$\sim$20 kpc,][]{larson16},
MCG+04-48-002 \citep[$\sim$25 kpc,][]{koss16},
TOL1238-364 \citep[$\sim$25 kpc,][]{temporin03},
and IC4518a \citep[$\sim$1 kpc,][]{bellocchi16}.
Two additional sources
(NGC~34 and ESO~286-IG019)
have a disturbed morphology,
sign of a past galactic interaction.
Moreover, some of the galaxies of this sample
(notably NGC~5128, 3C84 and 3C405)
are known to be part of groups or clusters,
so their morphology is unsettled
by probable continuous interactions
with nearby satellite galaxies.
Same as for the (U)LIRGs, 
interacting galaxies and systems 
with disturbed morphologies 
are typically characterized 
by higher molecular gas content 
and star-formation activity 
than isolated galaxies 
that may be due to tidal torques 
able to produce gas infall from the surrounding regions
\citep[e.g.][]{combes94, casasola04, pan18, moreno19}.

\section{Data collection}
\label{sec:data}

\begin{table*}
\centering
\caption{CO SLED transitions in units of 
$\log (L / L_{\odot})$}.
\label{tab:cosled}
\begin{threeparttable}

\begin{tabular}{lrrrrrrrrrrrrr}
\toprule
CO transition & 1--0 & 2--1 & 3--2 & 4--3 & 5--4 & 6--5 & 7--6 & 8--7 & 9--8 & 10--9 & 11--10 & 12--11 & 13--12 \\
Name            &       &       &        &        &         &        &        &        &        &         &        &         &         \\
\midrule
NGC0034         & 5.22 & 5.83 &  -- & <6.26 & 6.57 & 6.67 & 6.72 & 6.75 & 6.72 &  6.57 &  6.63 &  6.48 &  6.37 \\
UGC00545        & 5.42$^{\text{a}}$ & 6.25$^{\text{a}}$ & 6.92$^{\text{a}}$ &  -- &  -- & <7.08 & <7.01 & <7.15 & 7.22 &  <7.14 &  <7.18 &  <7.04 &  <7.17 \\
NGC1068         & 5.39$^{\text{b}}$ & 5.62$^{\text{b}}$ & 6.20$^{\text{a}}$ & 6.28 & 6.27 & 6.28 & 6.24 & 6.24 & 6.17 &  6.15 &  6.12 &  6.08 &  5.83 \\
3C84            & 4.85$^{\text{c}}$ & 4.48$^{\text{d}}$ & 5.92$^{\text{e}}$ & <6.48 & 6.39 & 6.32 & 6.25 & 6.33 & 6.41 &  6.45 &  6.32 &  6.31 &  6.13 \\
NGC1365         & 5.35 & 5.50 & 5.96 & 6.53 & 6.60 & 6.58 & 6.54 & 6.48 & 6.30 &  6.14 &  6.08 &  5.86 &  <5.77 \\
IRASF05189-2524 & 5.28$^{\text{a}}$ & 6.02$^{\text{a}}$ & 6.49$^{\text{a}}$ & <7.04$^{\text{a}}$ & 7.06 & 7.11 & 7.14 & 7.22 & 7.04 &  7.23 &  7.15 &  7.09 &  7.06 \\
IRAS07598+6508  & 5.78$^{\text{f}}$ & 6.57$^{\text{f}}$ &  -- &  -- & <8.08 & <7.70 & <7.77 & <7.62 &  -- &  <8.06 &  <8.00 &  <8.05 &  <8.02 \\
UGC05101        & 5.38$^{\text{a}}$ & 6.37$^{\text{a}}$ & 6.78$^{\text{a}}$ &  -- & 7.00 & 7.10 & 6.95 & 7.02 & 6.89 &  7.05 &  6.87 &  6.36 &  6.69 \\
NGC3227         & 4.15$^{\text{g}}$ & 4.82$^{\text{h}}$ & 5.23$^{\text{h}}$ & 5.41 & 5.48 & 5.44 & 5.30 & 5.34 & 5.19 &  5.11 &  5.24 &  5.15 &  <5.25 \\
NGC4151         & 2.55$^{\text{i}}$ & 3.23$^{\text{j}}$ &  -- & <5.14 &  -- & <4.84 & 4.66 & <5.02 & <5.14 &  5.26 &  <5.24 &  <5.18 &  5.03 \\
NGC4388         & 4.40$^{\text{h}}$ & 5.15$^{\text{h}}$ & 5.16$^{\text{h}}$ & 6.05 & 5.91 & 5.94 & 5.84 & 5.83 & 5.78 &  5.71 &  <5.96 &  <5.93 &  <5.90 \\
TOL1238-364     & 4.18$^{\text{k}}$ & 5.15$^{\text{k}}$ &  -- & <5.92$^{\text{k}}$ & 5.79 & 5.49 & 5.30 & 5.58 & 5.90 &  <5.98 &  <6.06 &  <5.90 &  <6.16 \\
Mrk0231         & 5.54$^{\text{a}}$ & 6.39$^{\text{a}}$ & 6.83$^{\text{a}}$ & 7.25$^{\text{a}}$ & 7.28 & 7.33 & 7.41 & 7.44 & 7.35 &  7.45 &  7.36 &  7.29 &  7.23 \\
MCG-03-34-064   &  -- &  -- &  -- & <6.22 & <6.20 & 5.97 & 5.96 & <6.25 & <6.31 &  6.38 &  6.05 &  6.09 &  6.14 \\
NGC5128         & 4.85$^{\text{l}}$ & 4.57$^{\text{m}}$ & 4.90$^{\text{m}}$ & 4.51 & 4.57 & 4.48 & 4.32 & 4.29 & <4.48 &  <4.27 &  <4.19 &  <4.24 &  <4.62 \\
NGC5135         & 5.19$^{\text{a}}$ & 6.00$^{\text{a}}$ & 6.38$^{\text{a}}$ & 6.51 & 6.61 & 6.61 & 6.49 & 6.37 & 6.31 &  6.13 &  6.03 &  5.95 &  5.65 \\
Mrk0463         & 5.12$^{\text{n}}$ & 5.08$^{\text{o}}$ &  -- &  -- & <7.05 & 6.81 & 6.67 & 6.61 & <7.03 &  6.37 &  <7.05 &  <7.08 &   -- \\
IC4518a         &  -- &  -- &  -- & 6.66 & 6.28 & 6.24 & 5.99 & 6.14 & <6.29 &  <6.16 &  6.25 &  <6.07 &  <6.28 \\
VV705           & 5.61$^{\text{a}}$ & 5.78$^{\text{p}}$ & 6.59$^{\text{a}}$ &  -- & 7.04 & 6.83 & 6.95 & 6.89 & <7.04 &  6.77 &  6.79 &  6.79 &  6.70 \\
PKS1549-79      &  -- &  -- &  -- &  -- & <8.26 & <7.95 & <7.71 &  -- &  -- &  <7.92 &  <7.81 &  <7.98 &  <7.99 \\
PG1613+658      & 5.49$^{\text{f}}$ &  -- &  -- &  -- & <7.99 & <8.00 &  -- & <7.59 &  -- &  <7.83 &   -- &  <7.94 &  <7.87 \\
NGC6240         & 5.63$^{\text{a}}$ & 6.59$^{\text{a}}$ & 7.10$^{\text{a}}$ & 7.46 & 7.59 & 7.69 & 7.75 & 7.78 & 7.75 &  7.72 &  7.65 &  7.59 &  7.52 \\
IRAS19254-7245  & 5.59$^{\text{n}}$ &  -- &  -- &  -- &  -- & 7.01 & 7.31 & 7.20 & 7.32 &  7.21 &  7.04 &  6.85 &  7.07 \\
3C405           & <4.12$^{\text{c}}$ &  -- &  -- &  -- & <7.21 & <7.01 & <6.85 &  -- & <7.21 &   -- &  <7.06 &   -- &  <7.13 \\
MCG+04-48-002   & 4.88$^{\text{q}}$ &  -- &  -- & <6.61 & 6.32 & 6.11 & 6.13 & 6.18 & <6.25 &  <6.33 &  <6.33 &  <6.18 &  <6.34 \\
IC5063          & 4.51$^{\text{h}}$ &  -- &  -- &  -- & <6.17 & <5.88 & 5.77 & <6.00 & <6.10 &  <6.15 &   -- &  <6.12 &  <6.17 \\
ESO286-IG019    & 5.50$^{\text{r}}$ &  -- & 6.30$^{\text{s}}$ &  -- & 7.22 & 7.13 & 7.30 & 7.36 & 7.22 &  7.37 &  7.33 &  7.25 &  7.18 \\
3C433           & <4.96$^{\text{c}}$ &  -- &  -- &  -- & <7.76 & <7.63 & <7.38 & <7.40 &  -- &  <7.37 &  <7.54 &  <7.55 &   -- \\
NGC7130         & 5.34$^{\text{q}}$ & 5.72$^{\text{p}}$ &  -- & 6.70 & 6.71 & 6.66 & 6.62 & 6.51 & 6.58 &  6.43 &  6.34 &  6.18 &  6.11 \\
NGC7172         & 4.75$^{\text{p}}$ & 5.25$^{\text{t}}$ &  -- & <6.10 & <5.79 & 5.64 & <5.41 & <5.62 & <5.59 &   -- &  <5.67 &   -- &  <5.77 \\
NGC7465         & 4.13$^{\text{g}}$ & 4.52$^{\text{u}}$ & 4.92$^{\text{e}}$ & <5.59 & 5.38 & <5.35 & <5.24 & <5.61 & <5.66 &  <5.64 &  <5.42 &   -- &   -- \\
NGC7469         & 5.24$^{\text{a}}$ & 6.02$^{\text{a}}$ & 6.44$^{\text{a}}$ & 6.69 & 6.83 & 6.80 & 6.71 & 6.62 & 6.58 &  6.40 &  6.35 &  6.20 &  6.15 \\
ESO148-IG002    & 5.29$^{\text{n}}$ &  -- &  -- &  -- & 6.99 & 7.04 & 7.15 & 7.13 & 7.14 &  7.09 &  7.02 &  6.89 &  7.03 \\
NGC7582         & 4.57$^{\text{h}}$ & 5.53$^{\text{t}}$ &  -- & 5.95 & 6.03 & 6.04 & 5.94 & 5.87 & 5.83 &  5.66 &  5.51 &  <5.41 &  <5.64 \\
NGC7674         & 5.70$^{\text{v}}$ & 5.93$^{\text{h}}$ & 6.26$^{\text{h}}$ & <6.95 & <6.57 & 6.32 & 6.09 & 6.36 & <6.68 &  <6.59 &  <6.63 &  6.59 &  <6.64 \\
\bottomrule
\end{tabular}
\begin{tablenotes}
\item \textbf{Notes.}
All the CO line luminosities are taken from
\cite{rosenberg15, mashian15, pearson16, kame16, lu17}
unless otherwise specified.
(\textit{a}) Data from \cite{papadopoulos12}: CO(1--0) was observed with
IRAM-30m (FWHM: $22"$), CO(2--1) (FWHM: $20"$),
CO(3--2) (FWHM: $14"$) and CO(4--3) (FWHM: $11"$) with JCMT.
(\textit{b}) Data from \cite{curran01}; 
(\textit{c}) Data from \cite{evans05}: 3C84 and 3C433 were observed with NRAO-12m
(FWHM: $55"$), 3C405 was observed with IRAM-30m (FWHM: $22"$).
(\textit{d}) Data from \cite{salome11}, observed with IRAM-30m (FWHM: $11"$).
(\textit{e}) Data from \cite{mao10}, observed with HHT (FWHM: $22"$).
(\textit{f}) Data from \cite{xia12}: CO(1--0) (FWHM: 22\farcs) and CO(2--1) (FWHM: $11"$)
were observed with IRAM-30m.
(\textit{g}) Data from \cite{maiolino97}, observed with NRAO-12m (FWHM: $55"$).
(\textit{h}) Data from \cite{israel20}; 
(\textit{i}) Data from \cite{dumas10}; 
(\textit{j}) Data from \cite{rigopoulou97}, observed with JCMT (FWHM: $20"$).
(\textit{k}) Data from \cite{pereira13}; 
(\textit{l}) Data from \cite{espada19}; 
(\textit{m}) Data from \cite{israel92}, observed with SEST (FWHM: $23"$),
CO(3--2) was observed with CSO (FWHM: $20"$).
(\textit{n}) Data from \cite{gao99}: ESO148-IG002 and
IRAS19254-7245 were observed with SEST (FWHM: $44"$),
Mrk0463 was observed with IRAM-30m (FWHM: $24"$).
(\textit{o}) Data from \cite{alloin92},
observed with IRAM-30m (FWHM: $13"$).
(\textit{p}) Data from \cite{albrecht07}; 
(\textit{q}) Data from \cite{gao04}; 
(\textit{r}) Data from \cite{ueda14}; 
(\textit{s}) Data from \cite{imanishi17}; 
(\textit{t}) Data from \cite{rosario18}; 
(\textit{u}) Data from \cite{monje11}; 
(\textit{v}) Data from \cite{young95}; 
\end{tablenotes}
\end{threeparttable}
\end{table*}

\subsection{X-ray data}
\label{sec:xsample}

We collect the best X-ray data 
available for our sample,
namely the intrinsic 2--10 keV luminosity
($L_{\text{X}}$), the column density ($N_{\rm H}$) 
of the obscuring material, and the photon index $\Gamma$
\citep{reynolds97, osterbrock06, singh11}
of the X-ray spectrum.
To minimize both the contribution from
host galaxy X-ray emission processes such as X-ray binaries,
and the obscuration of the AGN \citep{hickox18},
we prioritize hard-X \textit{NuSTAR} \citep[3-78 keV,][]{nustar}
and \textit{Swift/BAT}
\citep[15-150 keV,][]{swift, bat1, bat2} observations.

The data are taken from 
\citet{ricci17},
\cite{marchesi19, lacaria19} and
Salvestrini~et~al.~(2022, in prep.).
When not available in these works,
we take the $L_{\text{X}}$ and $N_{\text{H}}$ derived
from \textit{XMM-Newton} 
in the 0.5--10 keV band by \cite{brightman11}.
In Table~\ref{tab:sample}
we list the data together with their
references.
The final sample has a median\footnote{
The errors on the medians presented in this paper
always refer to the 16th and the 84th percentile
of the data distribution.}
$\log L_{\text{X}} \text{ [erg s}^{-1} \text{]}= 
42.8_{-0.5}^{+0.8}$.

$L_{\text{X}}$ is the intrinsic (i.e. unobscured)
luminosity of the AGN, after taking into account the 
obscuration of the gas along the line of sight.
Obscuration of AGN radiation
is usually measured in terms of
column density ($N_{\text{H}}$),
and it originates from the immediate vicinity
of the accretion disk, in the form
of a compact 
($\sim$0.1--10 pc)
dusty torus \citep{ramosalmeida17}.
However, as pointed out by recent works
\citep[e.g.][]{buchner17, damato20},
the obscuring gas can also
be associated with the host galaxy
on larger 
($\sim$10~pc--1~kpc)
scales.
For our sample,
the median $N_{\text{H}}$ is
$\log (N_{\text{H}} / \text{cm}^{-2}) =
23.5_{-1.8}^{+0.7}$,
with 27 of them being type 2 AGN
(i.e. they have $N_{\text{H}} > 10^{22}$
cm$^{-2}$, \citealt{hickox18}),
and six Compton-thick AGN
($N_{\text{H}} \geq 1.5 \times 10^{24}$
cm$^{-2}$, \citealt{matt00, comastri04}).
Assuming that this gas is distributed
over a sphere of 250 pc 
radius\footnote{See Section~\ref{sec:almadata}
for a definition of this radius},
the average gas density is
$\log (n / \text{cm}^{-3}) = 
2.6^{+0.7}_{-1.7}$.

\subsection{Herschel CO data}
\label{sec:cosample}

In the local Universe, 
the mid-$J$ and high-$J$ CO transitions have been observed with the \textit{Herschel} Space Observatory \citep{herschel}.
In particular, the transitions from CO(4--3)
(CO(5--4) for galaxies with $D_{\text{L}}>150 \rm \, Mpc$)
to CO(13--12) have been observed with the 
Spectral and Photometric Imaging Receiver
(SPIRE) Fourier Transform Spectrometer (FTS)
instrument \citep{spire} aboard \textit{Herschel}.
The beam full width at half maximum (FWHM)
of the SPIRE-FTS \textit{Herschel} observations \citep{lu17}
ranges from 16\farcs6 
at 200~$\mu$m 
to 42\farcs8 
at 650~$\mu$m,
respectively corresponding to the rest-frame wavelengths
of CO(13--12) and CO(4--3).
The beam FWHMs correspond to physical scales in the range
$\sim$6--14 kpc at the median redshift 
$z=0.02$ of our sample.

We collect SPIRE data from
\cite{rosenberg15, mashian15, pearson16, kame16, lu17},
which altogether account for
CO fluxes from 226 galaxies. 
In Table~\ref{tab:cosled} we report
the CO fluxes used in this work and, 
in case of multiple observations, we adopt
the mean and the standard deviation 
of the observed fluxes
as fiducial values.

\subsection{ALMA ancillary data}
\label{sec:almadata}

\begin{figure*}
    \includegraphics[width=\textwidth]{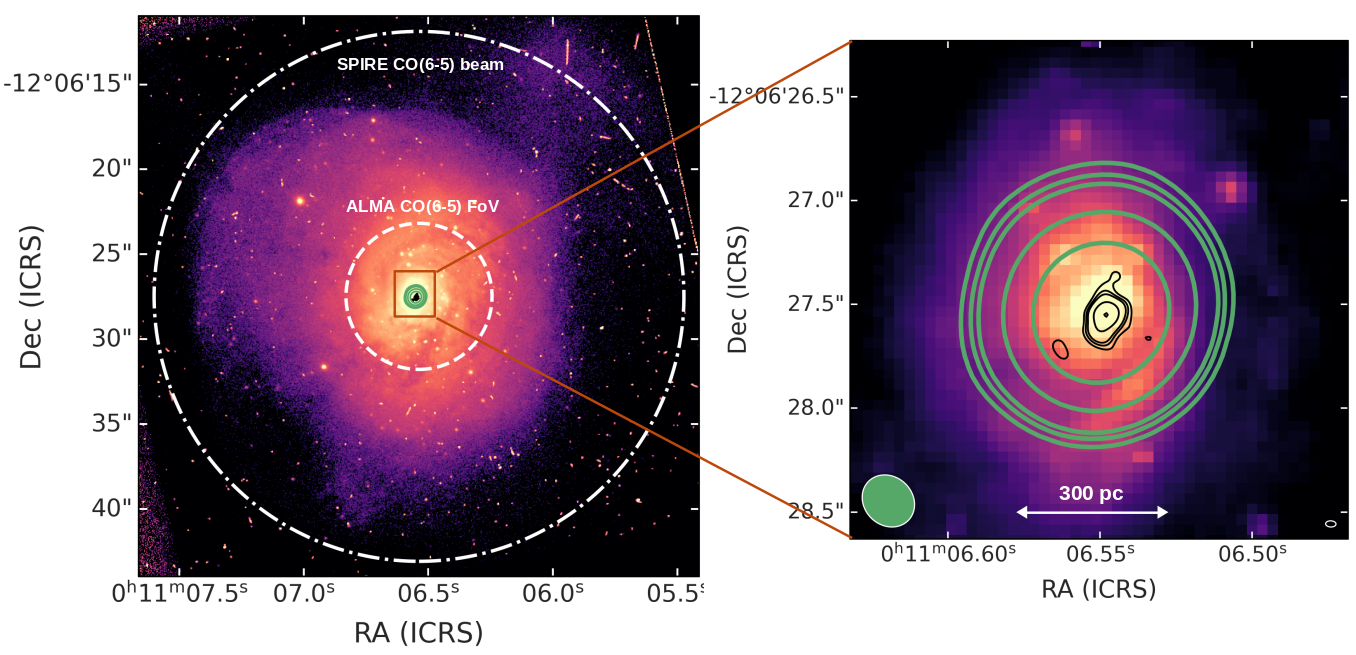}
    \caption{
    \textit{Left panel:} HST WFPC2 F606W image 
    of NGC~34 (from \protect\citealt{malkan98})
    with superimposed
    the contours of two ALMA CO(6--5) observations,
    in green at the resolution of 200~mas, in black of 35~mas.
    Both the contours are at the respective
    (3, 4, 5, 10, 20) $\times \, \sigma$,
    where $\sigma = 3.1$ Jy beam$^{-1}$ km s$^{-1}$ for the green lines
    and $\sigma = 0.27$ Jy beam$^{-1}$ km s$^{-1}$ for the black lines.
    The inner white dashed circle indicates the FoV
    of both ALMA observations, with a radius of 4\farcs3 ($\sim$1.7~kpc),
    while the outer dash-dotted circle represents
    the \textit{Herschel}/SPIRE-FTS beam FWHM
    for CO(6--5) observations,
    with a 15\farcs6 radius.
    \textit{Right panel:} zoom of the inner 1~kpc.
    Restored ALMA beams of the 200 and~35 mas
    images are shown as ellipses with white edges,
    at the bottom left (with the green area) 
    and right (with the black area), respectively.
    The 35~mas ALMA image has not been primary-beam corrected.
    }
    \label{fig:N34}
\end{figure*}

In local ($D \sim 1 \rm \, Mpc$) sources, 
the Atacama Large Millimeter Array (ALMA)
is able to resolve the morphology of CO emission 
at $\sim$100~pc scales,
from CO(1--0) to the mid-$J$ CO(6--5) line.
Higher-$J$ lines, which trace the dense/warm molecular gas
possibly influenced by the X-ray photons,
fall unfortunately out of the ALMA bands at low redshift.
From the ALMA 
archive\footnote{https://almascience.eso.org/asax/}
we therefore collect all the available maps 
of the highest possible CO transition -- namely the CO(6--5) -- for the galaxies in our sample.
We use these maps to infer the size of the high-density molecular gas region that cannot be estimated from the \textit{Herschel} data given their poor spatial resolution.
As the critical density of the CO transitions
increases with $J$
($n_{\text{crit}} \propto (J+1)^3$), and given that the gas density increases
as we get closer to the galaxy center, we expect
the higher-$J$ lines to originate from an area extended at most like CO(6--5) \citep[see e.g.][]{mingozzi18}.
We thus use the typical size of the CO(6--5) emitting region as an upper limit
for the AGN sphere of influence
on the molecular gas. 

Figure~\ref{fig:N34} shows --as an illustrative example--
the spatially resolved CO(6--5) emission from
NGC~34, a LIRG in our sample, hosting an obscured
($N_{\text{H}} = 10^{23.7}$ cm$^{-2}$) AGN
\citep{brightman11, mingozzi18}.
For this source, we retrieved two different
ALMA observations,
2011.0.00182.S (PI: Xu) and
2016.1.01223.S (PI: Baba),
both carried out in Band 9,
where the field of view (FoV)
is $\sim$ 8\farcs6,
but with different spatial resolutions
(200 and 35 mas, respectively) and maximum recoverable scales
($2''$ and $0.5''$, respectively). 
These scales correspond to
800 and 200 pc at the NGC~34 distance. 
The total flux of the CO(6--5) detection
with a resolution of 200 mas
is $S_{\text{CO(6--5)}} = 707 \pm 106$~Jy~km~s$^{-1}$,
obtained by \citet{mingozzi18}, using
CASA 4.5.2 \citep{mcmullin07} 
and a natural weighting scheme.
This flux, which is shown with the green contours 
in Figure~\ref{fig:N34}
\citep[see also][]{xu14, mingozzi18},
matches that recovered by \emph{Herschel}/SPIRE
within a much larger beam of 31\farcs2.
This means that
this ALMA observation, 
despite having a smaller FoV 
with respect to that of SPIRE, 
recovers all the CO(6--5) emission from the galaxy.

The high-resolution data 
(project ID 2016.1.01223.S, PI: Baba)
are plotted with black contours in Figure~\ref{fig:N34}
and have never been published so far.
We used the already calibrated and cleaned
data cube from the ALMA Archive.
For this data cube,
calibration and imaging have been done manually,
with a Briggs weighting (robust parameter of 0.5),
and passed the QA2 stage.
Using CASA 5.6 \citep{mcmullin07},
we produced the moment 0 map from the data cube
with the task \texttt{immoments}.
To estimate the flux, we performed a 2D Gaussian fit
with the task \texttt{imfit}, which returned
$62 \pm 3$ Jy km s$^{-1}$, less than
$ 10 \%$ of the total flux
measured by SPIRE-FTS ($920 \pm 56$ Jy km s$^{-1}$).
The reason for this discrepancy is that
this observation is limited by a much
smaller maximum recoverable scale, 
compared to the 200-mas data.
The emission consists of a single clump
of $r \lesssim 50$ pc.

In addition to NGC 34, 
we analysed ALMA CO(6--5) maps available for 
NGC~1068 \citep{burillo14},
IRAS~F05189--2524 (still unpublished),
NGC~5128 \citep{espada17}, 
NGC~5135 \citep{sabatini18}, 
NGC~6240 (still unpublished) 
and NGC~7130 \citep{zhao16}.
The images are shown in
Appendix~\ref{sec:CO6atlas}.
All these sources are characterized by 
spatially resolved CO(6--5) emission
arising from the galaxy center and extending up 
to $150-1000$ pc, with median $r=250$ pc.
We therefore assume that the bulk of 
higher-$J$ CO line luminosity -- 
for which we have only \emph{Herschel} at low resolution -- 
arise from a comparable region of radius $r=250$ pc. 
In what follows we use this size as an upper limit 
for $J \geq 6$ transitions emitting region.

\subsection{Dust continuum emission as a proxy for star formation}
\label{sec:dust}

Dust in active galaxies
can be heated by both the UV/optical photons
coming from black hole accretion,
and UV/optical photons 
associated to star-formation processes
\citep[e.g.][]{hatziminaoglou08, pozzi10, gruppioni16}.
In the first case, 
the dust is mostly circumnuclear,
which means it occupies
the central 100 pc at most
\citep[e.g.][]{hickox18};
in the second case the dust grains
reside in the star-forming regions
through the galaxy structure.
The emission of two dust components peaks 
at different infrared (IR) wavelengths,
due to the different temperatures:
the circumnuclear dust ($T\approx 60-100$ K)
peaks in the mid-IR, around $10 - 30 \, \mu$m
\citep{alonsoherrero11, feltre12}, 
while the galactic diffuse dust
is colder ($T\approx 20-30$ K), 
peaking in the far-IR
around $70 \sim 100 \, \mu$m \citep{dacunha08}.

For this reason we adopt the 
70 $\mu$m emission maps from
the \textit{Herschel} 
Photoconductor Array Camera and Spectrometer
\citep[PACS,][]{pacs}
as a proxy for SF 
in our sample galaxies. 
In this regime the AGN contamination, if any, 
accounts for a few percent,
and the spatial resolution at 70 $\mu$m
(FWHM = 5\farcs6,
corresponding to 
$\sim$0.17--13~kpc for our sample) 
is better than at longer wavelengths. 
We find suitable maps
for all the sources, except IRAS~07598+6508,
Mrk~463 and PKS~1549-79. We keep anyway these three
galaxies in our sample for completeness.

The 5\farcs6
spatial resolution
allows us to map
the distribution of SF,
assuming that all the 70 $\mu$m photons
trace the original stellar UV radiation.
From visual inspection,
SF is occurring
mostly in the central regions 
($r \sim 2$ kpc)
of our galaxies.
The procedure to extract the star formation rate (SFR)
and the radial profile of the Habing field 
from the 70 $\mu$m data
is outlined in Section~\ref{sec:pdr}.

\subsection{Low-J CO data}
\label{sec:lowj}

To complete the CO SLEDs observations from \textit{Herschel} 
discussed in Section~\ref{sec:cosample},
we collect (see Table~\ref{tab:cosled}) 
the low-$J$ fluxes available in the literature,
from CO(1--0) to CO(3--2).
These transitions have been observed
using several single-dish telescopes: 
the 14-m Five College Radio Astronomy Observatory (FCRAO),
the 15-m Swedish-ESO Submillimeter Telescope (SEST),
the 30-m Institut de Radioastronomie Millimétrique
Pico Veleta telescope (IRAM-30m),
the 12-m Atacama Pathfinder Experiment (APEX),
and the 15-m James Clerk Maxwell Telescope (JCMT).

We expect these low-$J$ CO lines to trace a larger area
than mid-$J$ and high-$J$ lines,
since they are characterized by lower $n_{\text{crit}}$ 
and lower excitation temperatures.
CO(1--0) is especially important since 
its flux is the most widely used proxy 
for the total molecular gas mass of a galaxy
\citep{bolatto13}.
For the closest galaxies,
their projection on the sky
could result larger than the 
telescope collecting area.
For this reason, when 
multiple observations are available,
we prioritize mosaics and larger beams.

Many authors have found that CO(1--0)
emitting gas has a exponential radial profile,
and that there is a relation between
the CO(1--0) scale length $r_{\text{CO}}$ 
and the optical radius $r_{25}$
\citep{leroy08, schruba11, villanueva21}.
Since the $\sim \! 30 \%$ of our sample contains 
highly inclined galaxies ($i \geq 60^{\circ}$), 
we follow \cite{boselli14} and \cite{casasola20}
assuming that the CO(1--0) emission is well described 
by an exponential decline both along 
the radius $r$ and above the galactic plane 
on the $z$ direction (3D method):
\begin{equation}
\label{eq:COprofile}
S_{\text{CO}}(r,z) = S_{\text{CO,tot}}
e^{-r/r_{\text{CO}}} e^{-|z|/z_{\text{CO}}} \, ,
\end{equation}
where $r_{\text{CO}} = 0.17 \, r_{25}$ 
and $z_{\text{CO}} = 0.01 \, r_{25}$,
as in \cite{casasola17} and \cite{boselli14}.
We stress that for galaxies with low inclination, 
the 3D method is analogous to the standard 2D approach, 
such as that developed by \cite{lisenfeld11}.
The adopted approach provides a median
$r_{\text{CO}} = 3.07^{+2.06}_{-1.48}$ kpc
for our sample.

\section{CO emission on global galactic scales}
\label{sec:SK}

Before investigating the PDR vs. XDR contribution to the molecular gas heating
in the center of our sample galaxies,
we want to see if, on the scale of the whole galaxy,
it is already possible to see the influence of the AGN
on the molecular gas phase.
We check how our active galaxies compare
to other active and non-active samples on the
Schmidt--Kennicutt plane \citep{schmidt59, k98}, 
which links the molecular gas surface density
$\Sigma_{\text{mol}}$ and the SFR
surface density $\Sigma_{\text{SFR}}$,
i.e. the star formation to its fuel.

We calculate the surface densities
$\Sigma_{\text{mol}}$ and $\Sigma_{\text{SFR}}$
within the CO radius r$_{\text{CO}}$,
defined as a fraction of the optical radius $r_{25}$
(see Section~\ref{sec:lowj}).
We derive the molecular mass from the CO(1--0) flux
in the following way.
For each source, we have the CO(1--0) flux $S_{\text{CO}}$,
measured within the telescope beam, with FWHM $2 \theta$,
in angular units (the factor 2 is due to the fact 
that the FWHM is a diameter, while we want a radius).
In spatial units (e.g. in pc) in the source reference frame,
this corresponds to a radius $r_{\theta}$,
so that the flux recovered by the telescope is:
\begin{equation}\label{eq:SCOrtheta}
S_{\text{CO}}(r_{\theta}) = \iint_{0}^{r_{\theta}}
S_{\text{CO}} dr dz = 
S_{\text{CO,tot}} (e^{-r_{\theta}/r_{\text{CO}}} - 1)
(e^{-r_{\theta}/z_{\text{CO}}} - 1) .
\end{equation}
If we put $r_{\text{CO}}$ instead of $r_{\theta}$
in Equation~\ref{eq:SCOrtheta}, we obtain that
$S_{\text{CO}}(r_{\text{CO}}) \approx 0.63 S_{\text{CO,tot}}$.
Given that we know $S_{\text{CO}}(r_{\theta})$
from observations, we can calculate the CO(1--0) flux
within $r_{\text{CO}}$:
\begin{equation}
\label{eq:COtot}
\begin{aligned}
S_{\text{CO}}(r_{\text{CO}}) = \iint_{0}^{r_{\text{CO}}}
S_{\text{CO}}(r,z) dr dz
= \frac{0.63 \, S_{\text{CO}}(r_{\theta})}
{(e^{-r_{\theta}/r_{\text{CO}}} - 1)
(e^{-r_{\theta}/z_{\text{CO}}} - 1)}
\end{aligned}
\end{equation}
We find a median ratio
$S_{\text{CO}}(r_{\text{CO}}) / S_{\text{CO}}(r_{\theta}) = 
0.70^{+0.30}_{-0.06}$, with only one galaxy
(NGC 5128) having 
$S_{\text{CO}}(r_{\text{CO}}) / 
S_{\text{CO}}(r_{\theta}) > 2$.
From the CO(1--0) flux
calculated within $r_{\text{CO}}$,
we estimate the molecular mass 
by using the following equation
from \cite{bolatto13}:
\begin{equation}
\label{eq:Mmol}
M_{\text{mol}} = 1.05 \times 10^{-16} X_{\text{CO}}
\frac{S_{\text{CO}} D^2_{\text{L}}}{1+z} \, M_{\odot} \, ,
\end{equation}
where $S_{\text{CO}}$ is 
the CO(1--0) flux in Jy km s$^{-1}$,
$D_{\text{L}}$ is the luminosity distance in Mpc,
$z$ is the redshift,
and $X_{\text{CO}}$ is the
CO-to-H$_2$ conversion factor.
The masses thus calculated
already take into account
the contribution of helium
and heavy elements.
To line up with the other samples included in our comparison,
we adopt a Milky Way value of
$X_{\text{CO}} = 2 \times 10^{20}$ cm$^{-2}$ 
(K km s$^{-1}$)$^{-1}$, corresponding to 
$\alpha_{\text{CO}} = 4.3$ M$_{\odot}$
(K km s$^{-1}$ pc$^{-2}$)$^{-1}$,
defined as the mass-to-light ratio
between $M_{\text{mol}}$ and 
the CO(1--0) luminosity.

We find $M_{\text{mol}}$ between
$10^{7.4}$ and $10^{10.6}$ M$_{\odot}$,
with median 
$\log (M_{\text{mol}} / M_{\odot}) = 10.1^{+0.3}_{-0.7}$.
These $M_{\text{mol}}$ are calculated
within $r_{\text{CO}}$: to extrapolate the results
to the whole galaxy ($r \rightarrow + \infty$),
a multiplicative factor of $1/0.63$ is needed.
The molecular masses calculated 
using Equations~\ref{eq:COtot}
and \ref{eq:Mmol}
are reported in Table~\ref{tab:sample},
while the uncorrected 
(i.e. the observed) CO luminosities
are the ones in Table~\ref{tab:cosled}.
We note that these masses could be upper limits,
since we are adopting a Milky Way value of 
$\alpha_{\text{CO}}$,
while it is thought that dusty (U)LIRGs and starburst
galaxies have a lower $\alpha_{\text{CO}} \approx 0.8$
M$_{\odot}$ (K km s$^{-1}$ pc$^{-2}$)$^{-1}$
\citep{downes98, bolatto13}.

The SFRs are estimated from
the radial profile $F_{70} (r)$ of the 70 $\mu$m 
photometry maps:
$\log \text{SFR} = \log L_{70} - 43.23$
\citep{calzetti10, ke12},
where $L_{70}$ is in units of erg s$^{-1}$ and
comes from the integration of
$F_{70} (r)$ up to r$_{\text{CO}}$.
This SFR calibration depends on the quantity 
of dust (it works better for dusty starburst galaxies)
and the stellar population mix, and works
better for galaxies with
$L_{70} > 4.4 \times 10^9$ L$_{\odot}$
\citep{calzetti10},
which is satisfied by the $\sim 90 \%$
of our galaxies.
Using this SFR calibration,
we find a median SFR $= 12.5^{+34.9}_{-9.8}$ 
M$_{\odot}$ yr$^{-1}$.

\begin{figure}
	\includegraphics[width=\columnwidth]{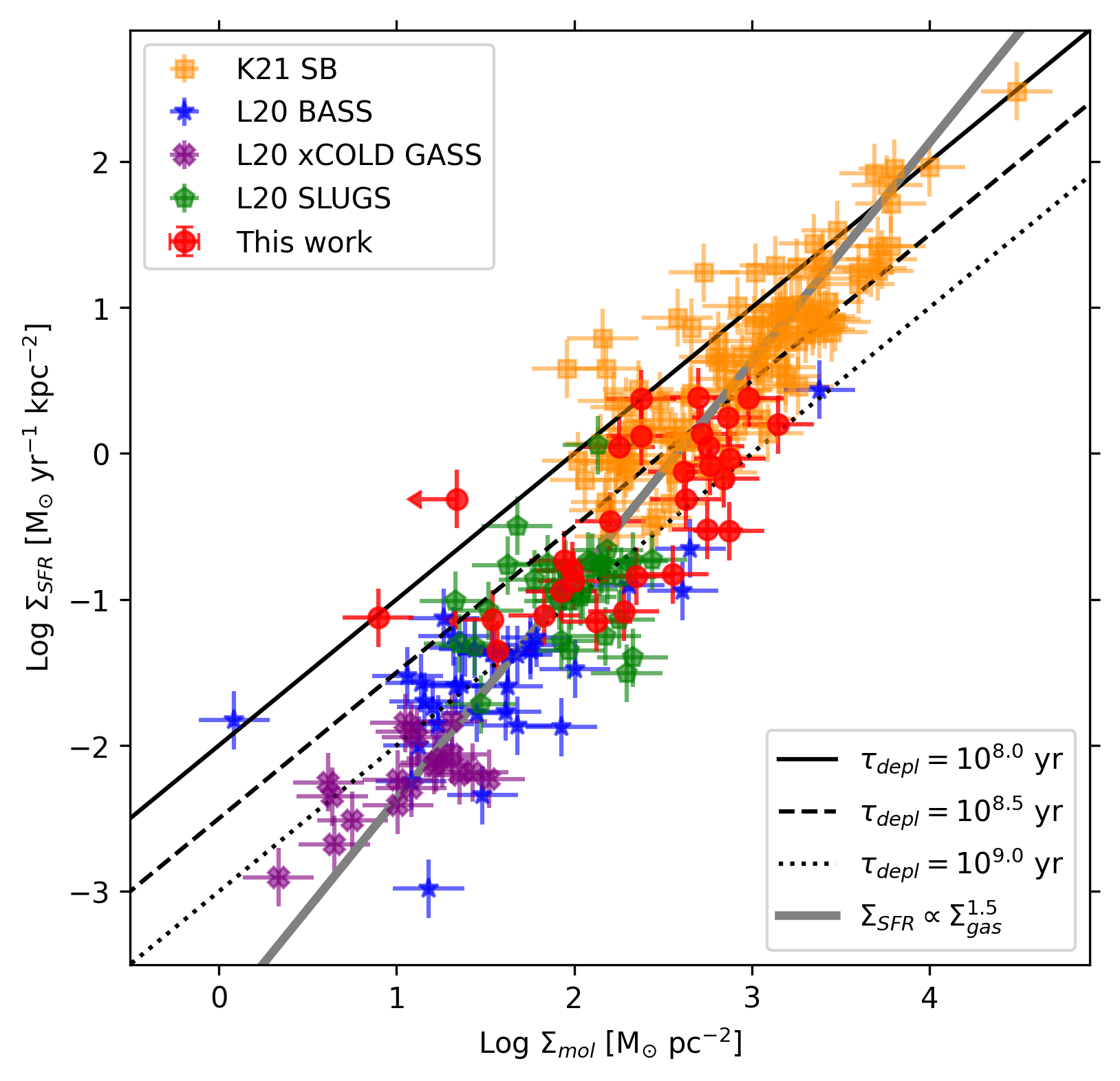}
    \caption{Schmidt--Kennicutt relation for our sample
    of active galaxies (red circles), the starburst
    sample from \protect\citealt{k21} (orange squares),
    and the AGN sample (blue stars) and normal SFG
    (pink crosses and green pentagons) from
    \protect\cite{lamperti20}.
    Lines of constant molecular gas depletion times
    are overlayed to the data.
    The gray solid line is the best fit for a single
    relation as reported by \protect\cite{k21}, namely 
    $\log \Sigma_{\text{SFR}} = 
    1.5 \log \Sigma_{\text{mol}} - 3.87$.
    All molecular surface densities were derived using
    the Milky Way value 
    $\alpha_{\text{CO}} = 4.3$ M$_{\odot}$
    (K km s$^{-1}$ pc$^{-2}$)$^{-1}$.}
    \label{fig:SK}
\end{figure}

In Figure~\ref{fig:SK}, 
we show our galaxies in the 
$\Sigma_{\text{mol}}$ -- $\Sigma_{\text{SFR}}$
plane, comparing them with starburst (SB) galaxies from
\citealt{k21} (K21, hereafter),  
AGN observed with Swift/BAT 
from the BASS sample \citep{ricci17},
star-forming galaxies (SFG) from the xCOLD GASS survey
\citep{saintonge17}, 
and IR luminous galaxies from SLUGS \citep{dunne00}. 
The latter three samples were gathered by 
\citealt{lamperti20} (L20, hereafter).

Our estimates of
$\Sigma_{\text{mol}}$ and $\Sigma_{\text{SFR}}$
mainly depend on the assumed CO exponential profile
and the SFR--70 $\mu$m calibration.
Following K21, we assign a conservative error
of $\pm 20 \%$ to both
$\Sigma_{\text{mol}}$ and $\Sigma_{\text{SFR}}$.
Since we could not recover the data errors
from every point of L20, we adopt the 
same $\pm 20 \%$ uncertainty 
also for their points.

We want to see if there is a difference
between normal SFGs and AGN on the 
$\Sigma_{\text{mol}} - \Sigma_{\text{SFR}}$ plane.
As shown in Figure~\ref{fig:SK},
our sample of AGN fit well in between
the starburst galaxies
of K21 and the mixed (AGN/SFGs)
sources from L20.
We note a gap between the K21 and L20 sources,
probably due to the difference in the 
area assumed for deriving the surface densities:
K21 calculate a circumnuclear starburst region
differently for every galaxy,
finding $r = 2.8^{+3.3}_{-1.2}$ kpc;
L20 instead use the CO observation 
beam area, which has a FWHM of 
15 arcsec for the SLUGS sample
and $20 \sim 22$ arcsec for both the xCOLD GASS
and the BASS sample 
(hence radii of $0.4 \sim 11$ kpc).
Overall, we find that, 
on the kpc-scale, an AGN effect
on the SF is not evident,
thus confirming earlier findings from
\cite{lamperti20}, and from
\cite{casasola15}, who studied
the Schmidt--Kennicutt relation for
four AGN from the NUGA sample
\citep{burillo03}.

In Fig.~\ref{fig:SK} we highlight 
the lines corresponding to constant depletion time,
$\tau_{\text{depl}} = 
\Sigma_{\text{mol}} / \Sigma_{\text{SFR}} = 
[10^8, 10^{8.5}, 10^9]$ yr, respectively.
For the galaxies in our sample, we find a median
$\log (\tau_{\text{depl}} / \text{yr}) = 8.9^{+0.4}_{-0.6}$,
similar to other studies of Seyferts
\citep[e.g.][]{salvestrini20}, and
slightly lower than typical values for
local inactive SFGs
\citep[][all find a median 
$\tau_{\text{depl}} \sim 2 \times 10^9$ yr]{bigiel08, utomo18, leroy21}.
Conversely, typical progenitors of ellipticals 
or proto-spheroids galaxy models
\citep{calura14} require 
$\tau_{\text{depl}} \sim 2 \times 10^7$ yr,
while dusty sub-millimeter galaxies (SMG),
which are mostly hyperluminous
infrared galaxies (HyLIRG, 
$L_{\text{IR}} \geq 10^{13}$ L$_{\odot}$)
at moderately high redshift ($z \sim 3$)
can have even shorter
$\tau_{\text{depl}} \leq 10^7$ yr
\citep{carilli13},
but these are probably extreme
and rare objects \citep{heckman14}.

From a classical evolutionary perspective,
active, interacting (U)LIRGs
are thought to be an intermediate
stage between a late-type
SFG and a quiescent early-type
galaxy \citep{hopkins08}.
From more recent works it seems
that interacting and merging systems
can account only for the formation
of the most massive ellipticals,
while slow secular processes
(in the local Universe)
or rapid instabilities in clumpy gaseous
disks (at high $z$)
are responsible for the evolution
of the bulk of the galaxies
\citep{heckman14}.
Within the limits of our analysis,
we do not see a strong effect
of AGN feedback on $\tau_{\text{depl}}$
at kpc-scales,
but that its impact also depends
on the choice of $\alpha_{\text{CO}}$.

\section{CO emission in the galaxy centers}
\label{sec:results}

\begin{figure*}
\centering
    \begin{subfigure}[b]{0.49\textwidth}
    \centering
    \includegraphics[width=\textwidth]
    {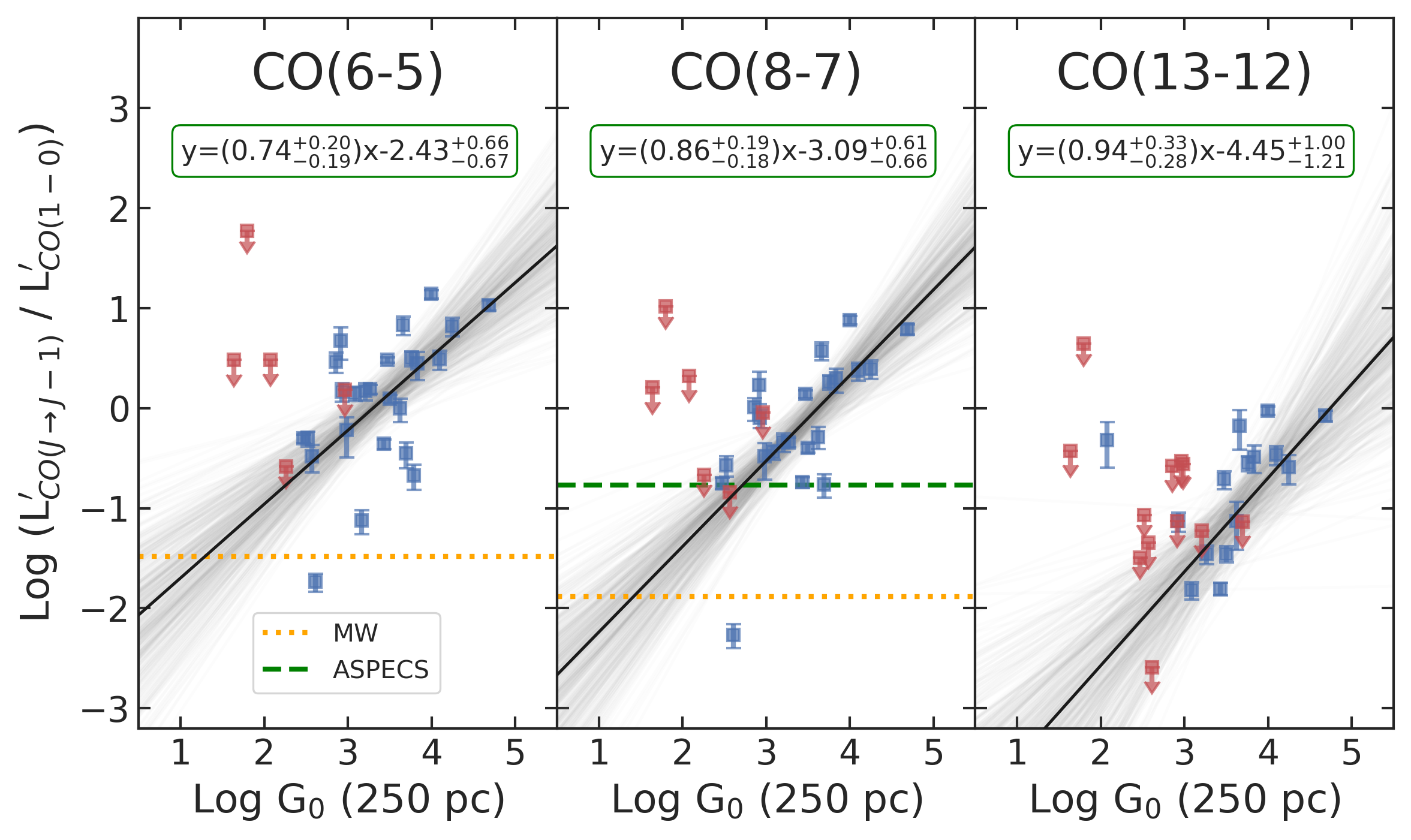}
    \caption{}
    \label{fig:G0CO1some}
    \end{subfigure}
\hfill
    \begin{subfigure}[b]{0.49\textwidth}
    \centering
    \includegraphics[width=\textwidth]
    {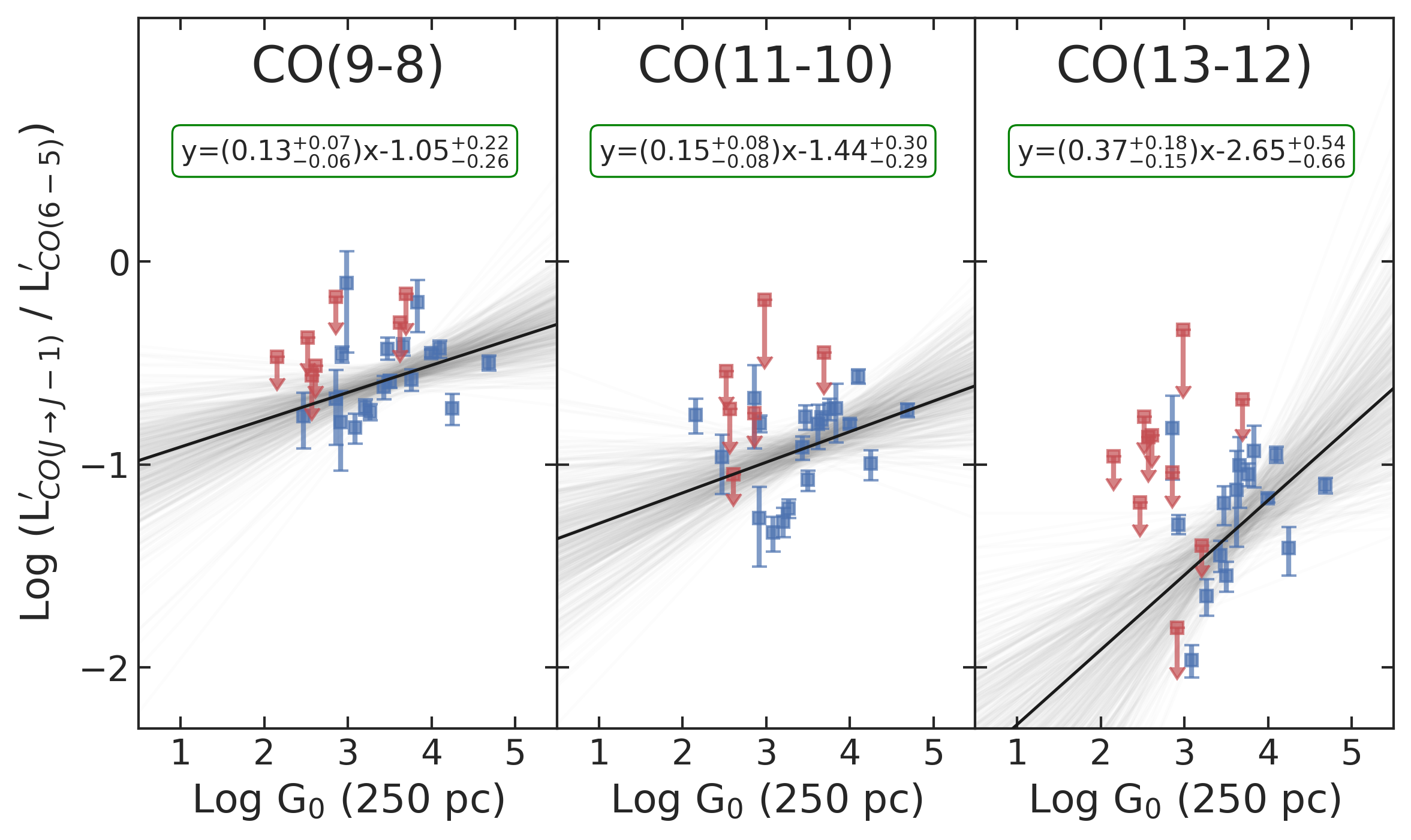}
    \caption{}
    \label{fig:G0CO6some}
    \end{subfigure}
\caption{CO line ratios as a function 
    of the Habing field, $G_0$,
    measured at $r=250$ pc 
    (see Section~\ref{sec:pdr}). 
    We consider both the luminosity ratios
    $L^{\prime}_{\text{CO}(J \rightarrow J-1)} / L^{\prime}_{\text{CO}(1 \rightarrow 0)}$
    with respect to the CO(1--0)
    (left panel, \ref{fig:G0CO1some}) and 
    $L^{\prime}_{\text{CO}(J \rightarrow J-1)} / L^{\prime}_{\text{CO}(6 \rightarrow 5)}$
    with respect to the CO(6--5)
    line luminosity (right panel, \ref{fig:G0CO6some}).
    The luminosities $L^{\prime}$ are in units of
    K km s$^{-1}$ pc$^{-2}$, and
    $J$ is indicated on the top of each panel.
    Blue squares indicate $3 \sigma$ detections, while
    red squares with downward arrow
    indicate $<3\sigma$ detections in the
    higher-$J$ line (i.e. censored data).
    The solid black line is the regression fit,
    with the underlying grey lines showing the fits drawn
    from the posterior distribution.
    When available, the Milky Way
    (dotted orange line, data from \citealt{fixsen99})
    and the ASPECS AGN
    (green dashed line, data from \citealt{boogaard20})
    CO ratios are also shown.
    }
\label{fig:G0COratios}
\end{figure*}

We now focus on the CO emission in the inner 500 pc 
(i.e. up to $r=250$ pc from the center) 
with the aim of assessing the 
relative contribution of PDR and/or XDR
to the molecular gas in the vicinity of the AGN.
To this goal, we exploit the line ratios with respect to 
CO(1--0) and CO(6--5):
$L^{\prime}_{\text{CO} (J \rightarrow J-1)} / 
L^{\prime}_{\text{CO} (1-0}$ 
(i.e. high-$J$/low-$J$ ratios) and
$L^{\prime}_{\text{CO} (J \rightarrow J-1)} / 
L^{\prime}_{\text{CO} (6-5)}$
(i.e. high-$J$/mid-$J$ ratios),
where all $L^{\prime}_{\text{CO}}$ are 
in units of K km s$^{-1}$ pc$^{-2}$.
We use the CO(1--0) 
theoretical profile (Equation~\ref{eq:COtot})
to calculate the flux within $r=250$ pc:
\begin{equation}
\label{eq:CO500}
S_{\text{CO}}(r \leq 250 \text{ pc}) = 
S_{\text{CO,tot}}
(e^{-250 \text{pc}/r_{\text{CO}}} - 1)
(e^{-250 \text{pc}/z_{\text{CO}}} - 1)
\end{equation}
Conversely, we do not correct the other CO lines:
we know (Section~\ref{sec:almadata}) that CO(6--5) emission
is mostly confined within the central 250 pc,
and the same should likely apply for higher-$J$ lines.
There are few studies that map the size 
of other low-$J$ lines than CO(1--0):
\cite{casasola15} compares
CO(1--0), CO(2--1) and CO(3--2) images for 4 nearby
active galaxies (none of which is part of this sample),
finding a similar physical size for the first two transitions
and a halved size (mean $\sim 500$ pc)
for the available CO(3--2) maps;
NGC 1068, however, has a CO(3--2) emission
which extends beyond the central 2 kpc \citep{burillo14}.
Among our sample of galaxies, 
\cite{dasyra16} have published a CO(4--3) image
of IC 5063, which has a similar size ($\sim 1$ kpc)
of its CO(2--1) emission.
CO(4--3) images of IRAS F05189--2524,
NGC 5135, ESO 286--IG019, NGC 7130, NGC 7469 and ESO 148--IG002,
among other (U)LIRGs, are published by
\cite{michiyama21}, who find emitting sizes
for the aforementioned galaxies
between 1 and 5 kpc.
Since these low-$J$ CO transitions are not
the focus of the present work, 
and since we do not have a theoretical
radial profile to correct them,
we leave them unaltered, 
and put the relative plots
only in the Appendix~\ref{sec:COratiofull}.

In the next two subsections, we derive
the fluxes of FUV and X-ray photons, 
which are
the heating drivers in PDRs and XDRs, respectively,
and we compare them
with the CO line ratios.

\subsection{PDR}
\label{sec:pdr}

The FUV flux (also often referred to as interstellar
radiation field) is measured in Habing units $G_0$,
where $G_0 = 1$ corresponds to 
its value in the solar neighbourhood:
$1.6 \times 10^{-3}$ erg cm$^{-2}$ s$^{-1}$
in the FUV band \citep{habing68}.
As discussed in Section~\ref{sec:dust},
the FUV photons are efficiently 
absorbed by dust grains,
which re-emit energy in the infrared (IR),
especially around 70$\mu$m
\citep[given typical dust temperatures;][]{dacunha08}.
Since our systems are powerful IR-emitters
(with median
$\log (L_{\text{IR}} / L_{\odot}) = 11.4^{+0.6}_{-0.9}$),
we assume that all the FUV photons
are processed by dust and re-emitted at 70 $\mu$m.

We use \textit{Herschel}/PACS 70 $\mu$m High Level 
Images\footnote{https://irsa.ipac.caltech.edu/data/Herschel/HHLI/overview.html}
to extract a value for $G_0$, 
assuming that all FUV photons are absorbed by dust grains
and re-emitted at 70 $\mu$m.
To do so, we fit the radial profile of the 
70 $\mu$m photometric map with a Sersic function:
\begin{equation}
F(R) = F_e \exp \left\{ -b_n \left[ \left( 
\frac{R}{R_e} \right)^{1/n} -1 \right] \right\} .
\end{equation}
The free parameters of this fit are $F_e$, $R_e$ and $n$,
while $b_n$ is a constant that depends on $n$
\citep{sersic63}.
We then divide the normalization flux $F_e$
by $1.6 \times 10^{-3}$ erg cm$^{-2}$ s$^{-1}$,
obtaining a profile in $G_0$ units.
In this way we find values corresponding to the radius $R_e$,
with median $\log G_0 (R_e) = 2.6^{+0.5}_{-0.8}$,
which is similar to what
\cite{farrah13} and \cite{diazsantos17} found
for local (U)LIRGs, in
the HERUS ($10^{2.2} < G_0 < 10^{3.6}$)
and the GOALS ($10^{1} < G_0 < 10^{3.5}$) samples,
respectively.
It is important to note that in these works,
as in most of the literature, $G_0$ is derived 
from PDR calculations fitting the observed line emission, thus relying on PDR codes
as e.g. the PDR Toolbox \citep{pound08}
and \textsc{Cloudy} \citep{ferland17}.
Here, instead, we observationally derive $G_0$
and we use the fitted profile to estimate its value
at different radii.
$G_0$ increases at smaller radii due to the
higher SFR in the circumnuclear region,
and the consequent high FUV irradiation.
At $r=250$ pc, we find a median
$\log G_0 (250 \text{pc}) = 3.1^{+0.7}_{-0.8}$.
We look then for correlations 
between the CO line ratios and $G_0$
(from now on when we refer to $G_0$ values
we mean measured at $r=250$ pc), 
to understand if the FUV irradiation can explain
by itself the observed CO emission at the center of
local active galaxies.

In Figure~\ref{fig:G0COratios} we show the 
CO(6--5)/CO(1--0), CO(8--7)/CO(1--0),
and CO(13--12)/CO(1--0) luminosity ratios on the left panel,
and the CO(9--8)/CO(6--5), CO(11--10)/CO(6--5) and
CO(13--12)/CO(6--5) ratios on the right panel,
as a function of $G_0$.
All the other CO line ratios are presented 
in the Appendix~\ref{sec:COratiofull}.
We see an overall trend,
for high-$G_0$ galaxies,
to show increasing 
high-$J$/low-$J$ and high-$J$/mid-$J$ ratios.

We fit a regression line
with the \texttt{Linmix} algorithm \citep{kelly},
which evaluates the likelihood in presence
of censored data (i.e. upper limits).
Linmix computes the likelihood function
by convolving multiple
(we use two, since adding more has
a negligible effect on our results)
hierarchical Gaussian distributions.
We also tried to fit only the detections
with an ordinary least squares regression
and with a bootstrapped version of the same algorithm,
finding limited differences 
with respect to the Linmix regression,
which includes the censored data.
Since an important fraction
(between 20 and 50 $\%$, depending on the transition)
of the high-$J$ CO fluxes are actually upper limits
(see Table~\ref{tab:cosled}),
we plot the Linmix results in
Figures~\ref{fig:G0COratios}~and~\ref{fig:FXCOratios}
and in Appendix~\ref{sec:COratiofull}.

\begin{figure*}
\centering
    \begin{subfigure}[b]{0.49\textwidth}
    \centering
    \includegraphics[width=\textwidth]
    {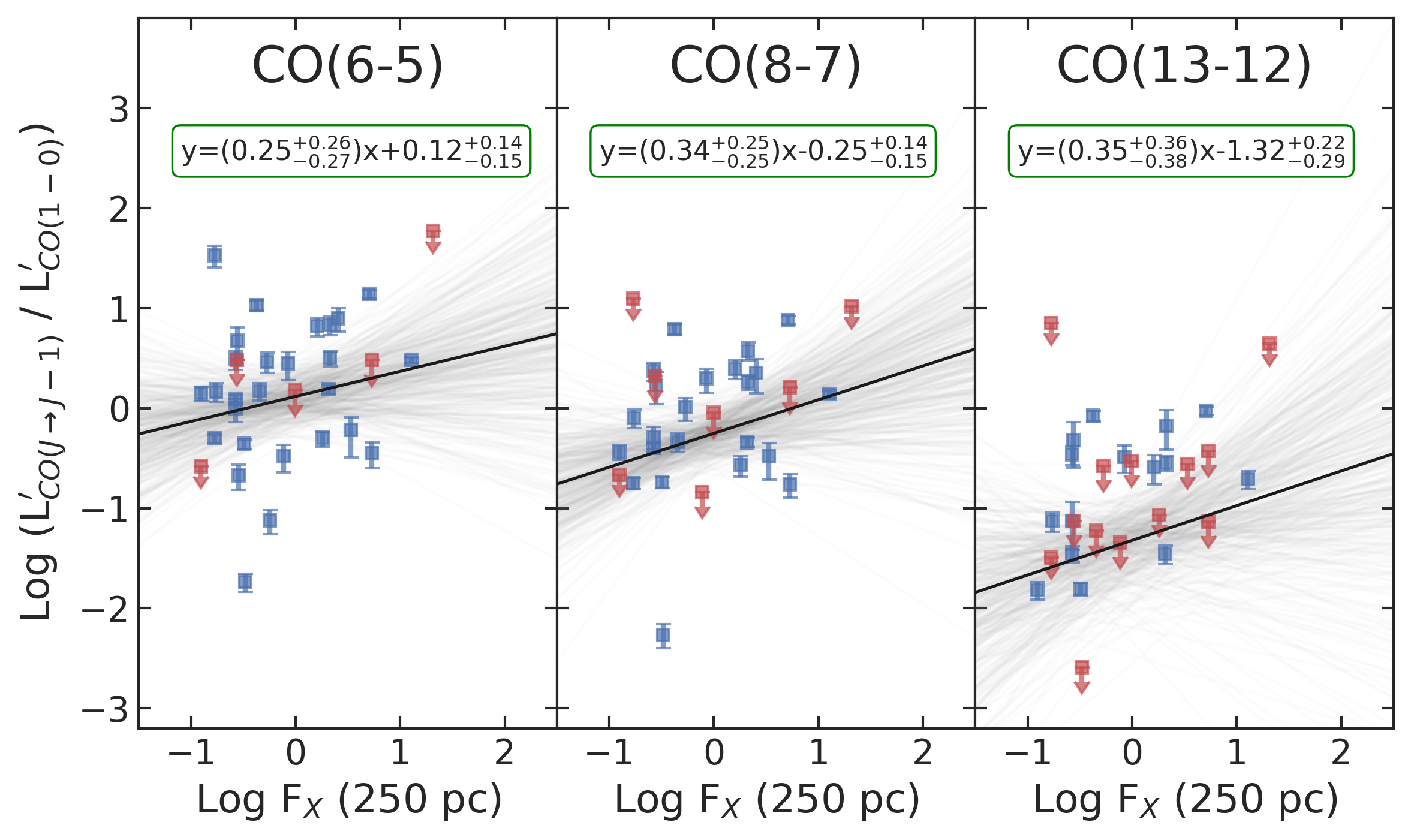}
    \caption{}
    \label{fig:FXCO1some}
    \end{subfigure}
\hfill
    \begin{subfigure}[b]{0.49\textwidth}
    \centering
    \includegraphics[width=\textwidth]
    {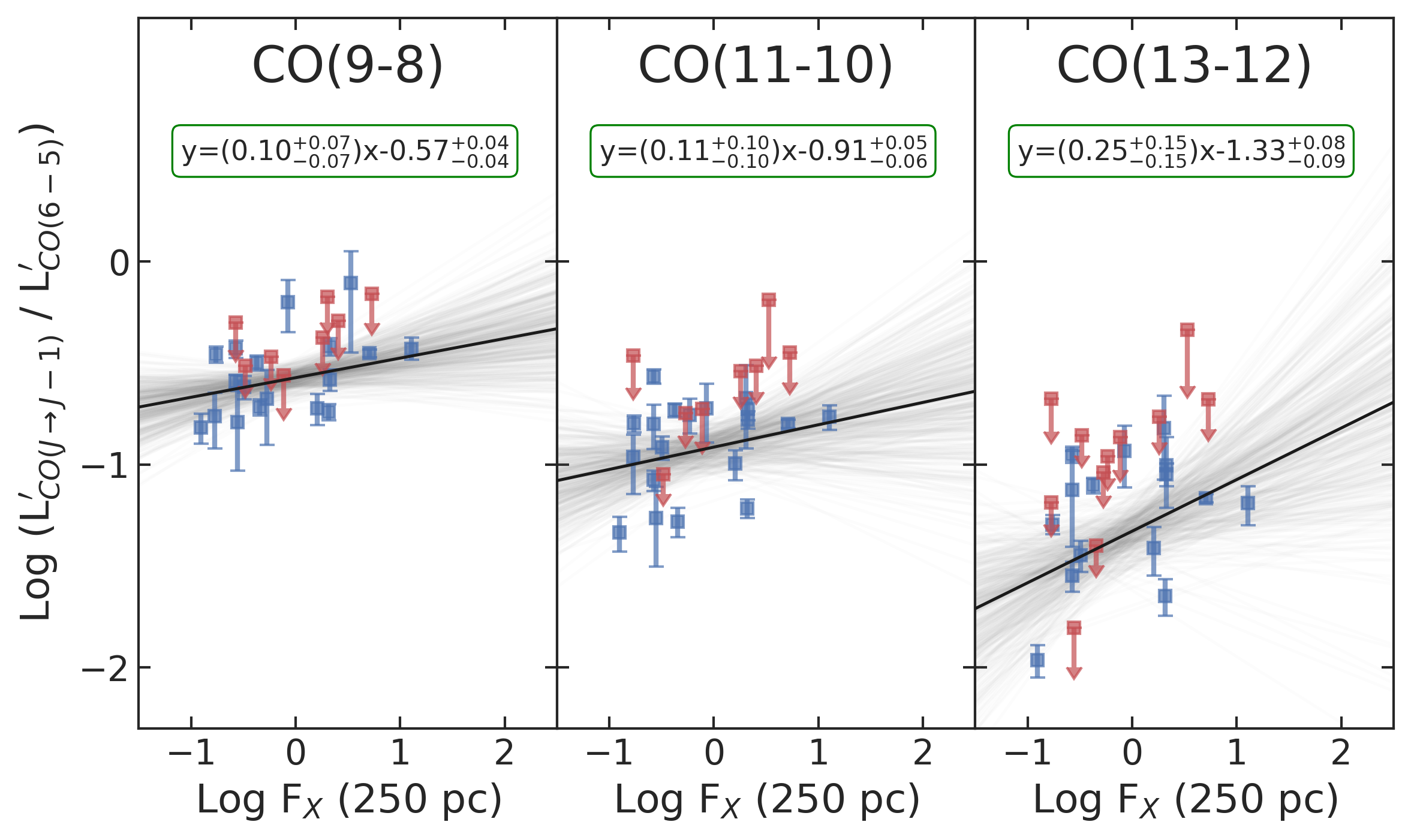}
    \caption{}
    \label{fig:FXCO6some}
    \end{subfigure}
\caption{CO ratios as a function of $F_{\text{X}}$,
    in units of erg s$^{-1}$ cm$^{-2}$,
    derived at $r=250$ pc
    (see Section~\ref{sec:xdr}).
    We consider both the luminosity ratios
    $L^{\prime}_{\text{CO}(J \rightarrow J-1)} / L^{\prime}_{\text{CO}(1 \rightarrow 0)}$
    with respect to the CO(1--0)
    (left panel, \ref{fig:FXCO1some}) and 
    $L^{\prime}_{\text{CO}(J \rightarrow J-1)} / L^{\prime}_{\text{CO}(6 \rightarrow 5)}$
    with respect to the CO(6--5)
    line luminosity (right panel, \ref{fig:FXCO6some}).
    The luminosities $L^{\prime}$ are in units of
    K km s$^{-1}$ pc$^{-2}$, and
    $J$ is indicated on the top of each panel.
    Blue squares indicate $3 \sigma$ detections
    in both lines;
    red squares with downward arrow
    indicate $<3 \sigma$ detections in the
    higher-$J$ line. The solid black line is the regression fit,
    with the underlying grey lines showing the fits drawn
    from the posterior distribution.}
\label{fig:FXCOratios}
\end{figure*}

We find steeper slopes for the CO(1--0) ratios,
and a trend of increasing steepness with $J$
for both ratios.
However, almost all the regression slopes return
a sub-linear relation between the 
CO line ratios and $G_0$, with slopes $0.3 - 1.1$ for the
CO(1--0) ratios, and $0.1 - 0.4$ for the CO(6--5)
ratios.
These findings suggest that the CO excitations 
are not strongly dependent on the radiative field $G_0$, 
and other excitation mechanisms may contribute 
to the CO line emission.

We also plot in Figure~\ref{fig:G0COratios}
the median line ratios for
the Milky Way \citep[MW,]{fixsen99} and
the AGN from the ASPECS \citep{walter16}
AGN sample \citep{boogaard20}.
The MW has a lower CO ratio than most of our
sources, which is expected since our galaxies 
are forming stars at a higher rate than the MW 
and host an AGN.
The ASPECS AGNs are instead bright
(L$_{\text{IR}} \sim 10^{12} L_{\odot}$)
and have a median CO ratio comparable to our active galaxies.
These AGN are located at $z \sim 1-3$,
at the peak of the cosmic SF history
\citep{madau14}.

\subsection{XDR}
\label{sec:xdr}

We use the $L_X$ and $N_H$ derived for our sample
(see Section~\ref{sec:xsample} for details) 
to estimate the unobscured X-ray flux, $F_X = L_X / (4 \pi r^2)$, 
illuminating the GMCs located at $r=250$ pc 
from the center of our galaxies.
We find a median
$\log (F_{\text{X}} / \text{erg s}^{-1} \text{cm}^{-2}))
= -0.1_{-0.5}^{+0.8}$.

According to theoretical \citep{kawakatu08}
and observational works \citep{davies07, esquej14, motter21},
the circumnuclear star-forming region
directly influenced by the AGN has a $\approx 100$ pc radius.
However,
with the available ALMA data (Section~\ref{sec:almadata})
we could study only up to the mid-$J$ CO(6--5) emission,
which is confined, on average, within a $\sim 250$ pc radius.
We, therefore, calculate our X-ray fluxes at this $r=250$ pc.
It is also possible to estimate $F_X$
from XDR numerical modelling, as done by
\cite{vanderwerf10, pozzi17, mingozzi18}.
Those works all find higher $F_{\text{X}}$
than ours for three galaxies of our sample
(respectively Mrk 231, NGC 7130, and NGC 34).
This may imply that
$r=250$ pc is a too large radius for the central XDR.

The X-ray flux $F_{\text{X}}$ does not account
for the obscuration of the X-ray photons before
they strike the molecular gas.
It is therefore useful to calculate
the local (i.e. accounting for the absorption) 
X-ray energy deposition rate
per particle $H_{\text{X}}$.
It can be estimated
from the following formula \citep{maloney96}:
\begin{equation}
\label{eq:HX}
H_{\text{X}} \approx 7 \times 10^{-22} 
L_{44} \, r_2^{-2} N_{22}^{-1}
\; \; \text{erg s}^{-1},
\end{equation}
where the X-ray luminosity is $L_X = 10^{44} L_{44}$
erg s$^{-1}$, the distance to the X-ray source
is $r = 10^2 r_2$ pc and the attenuating column
density is $N_H = 10^{22} N_{22}$ cm$^{-2}$.
We find a median
$\log (H_{\text{X}} / (\text{erg s}^{-1})
= -25.3^{+1.1}_{-0.9}$.
We use the $N_{\text{H}}$ measured from the X-ray
spectrum (Section~\ref{sec:xsample})
to estimate $H_{\text{X}}$.
Although a Compton-thick gas ($N_{\text{H}} > 10^{24}$
cm$^{-2}$) is generally associated to small-scale 
structures like a dusty molecular torus,
Compton-thin gas (as it is for $65 \%$ of our sample)
may be part of the same circumnuclear gas
we are studying from molecular and IR emission
\citep{ballantyne08, hickox18}.
In this case,
the $H_{\text{X}} \propto N_{\text{H}}^{-1}$
we calculate
from Equation~\ref{eq:HX} could be underestimated,
since there would be a lower $N_{\text{H}}$
between the XDR and the AGN.

A key physical quantity affecting 
the XDR emission, and
directly proportional to $H_{\text{X}}/n$,
is the effective ionization parameter,
defined \citep{maloney96, galliano03, motter21} as:
\begin{equation}
\xi_{\text{eff}} = 1.06 \times 10^{-2}
L_{44} \, r_2^{-2} \, N_{22}^{-\alpha} \, n_5^{-1}
\; \; \text{erg cm}^3 \text{ s}^{-1},
\end{equation}
where the density of the XDR gas is 
$n = 10^5 n_5$ cm$^{-3}$,
$\alpha = (\Gamma + 2/3)/(8/3)$ depends
on the photon index $\Gamma$ of the X-ray spectrum
\citep{kawamuro20}
and the other quantities are the same
defined above for $H_{\text{X}}$.
For a representative fixed value of 
$n_5 = 0.1$ we find a median
$\log \xi_{\text{eff}} / (\text{erg cm}^3
\text{s}^{-1}) = -4.2^{+1.9}_{-1.0}$.
These values are very low when compared
to the theoretical values found in \cite{maloney96}
models (e.g. their Figure 7)
and to the observed values found in \cite{motter21},
who calculated $\xi_{\text{eff}}$ 
for the active galaxy NGC 34,
also present in our sample.
\cite{motter21} used $N_{\text{H}}$
derived from radio observations
(which is 1 dex lower than the one we use 
for NGC 34, derived from X-rays),
and calculated $\xi_{\text{eff}}$ 
at distances from the AGN between 40 and 120 pc,
thus finding values $\sim \! 2$ dex higher
than us.
When taking into account these differences,
the results are compatible.
Again, this may be a clue that
at $r=250$ pc we cannot yet see the AGN impact.

In Figure~\ref{fig:FXCOratios} we plot
the same luminosity line ratios
(CO(6--5)/CO(1--0), CO(8--7)/CO(1--0) and
CO(13--12)/CO(1--0) on the left panel,
CO(9--8)/CO(6--5), CO(11--10)/CO(6--5) and
CO(13--12)/CO(6--5) on the right panel)
analysed in Figure~\ref{fig:G0COratios},
as a function of $F_{\text{X}}$ only, since
both $H_{\text{X}}$ and $\xi_{\text{eff}}$
were showing, compared to $F_{\text{X}}$,
less defined trends.
The other CO line ratios and their regression
fits, as function of $F_{\text{X}}$,
are presented in Appendix~\ref{sec:COratiofull}.

\begin{figure*}
\centering
    \begin{subfigure}[t]{0.49\textwidth}
    \centering
    \includegraphics[width=\textwidth]
    {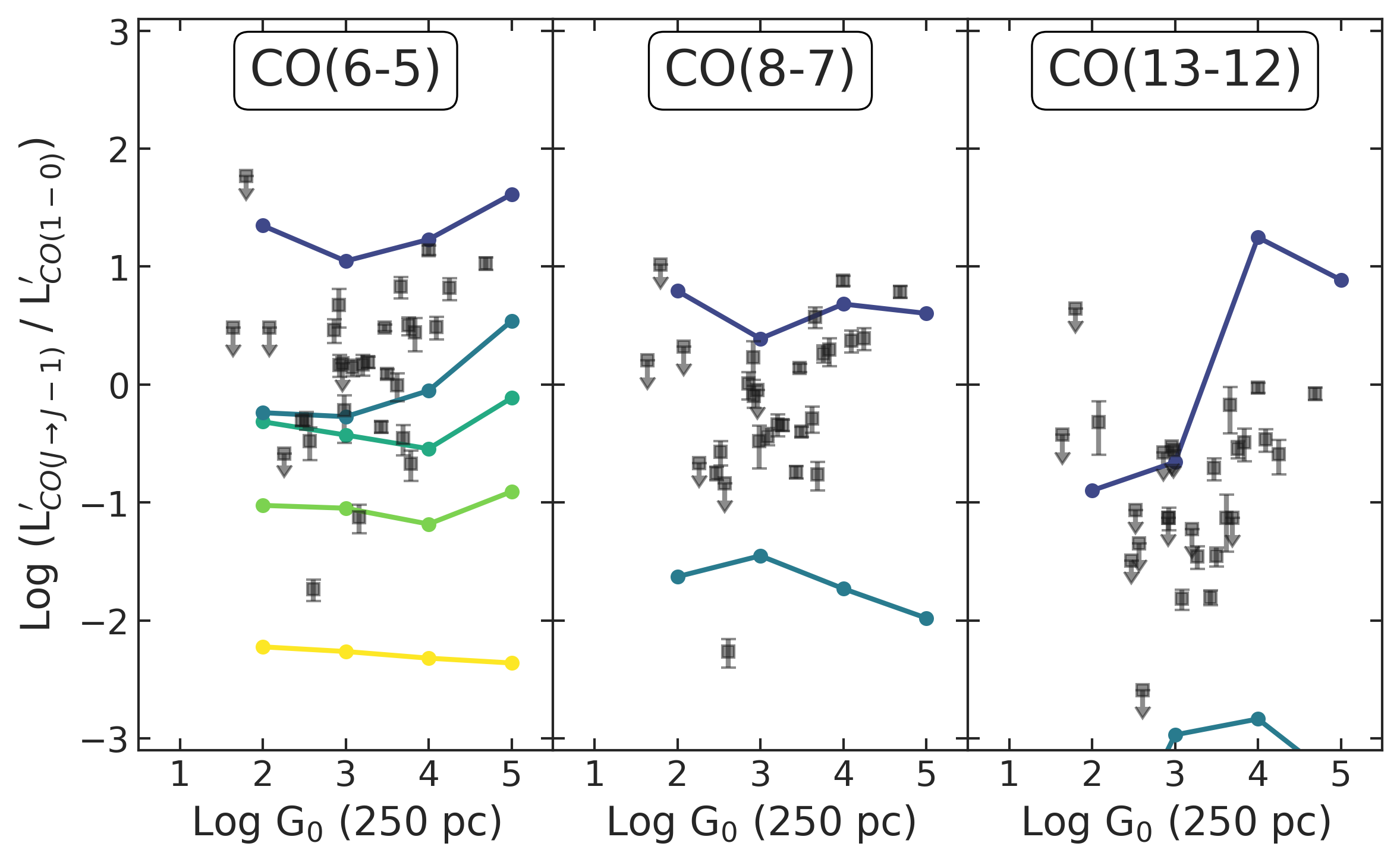}
    \end{subfigure}
\hfill
    \begin{subfigure}[t]{0.49\textwidth}
    \centering
    \includegraphics[width=\textwidth]
    {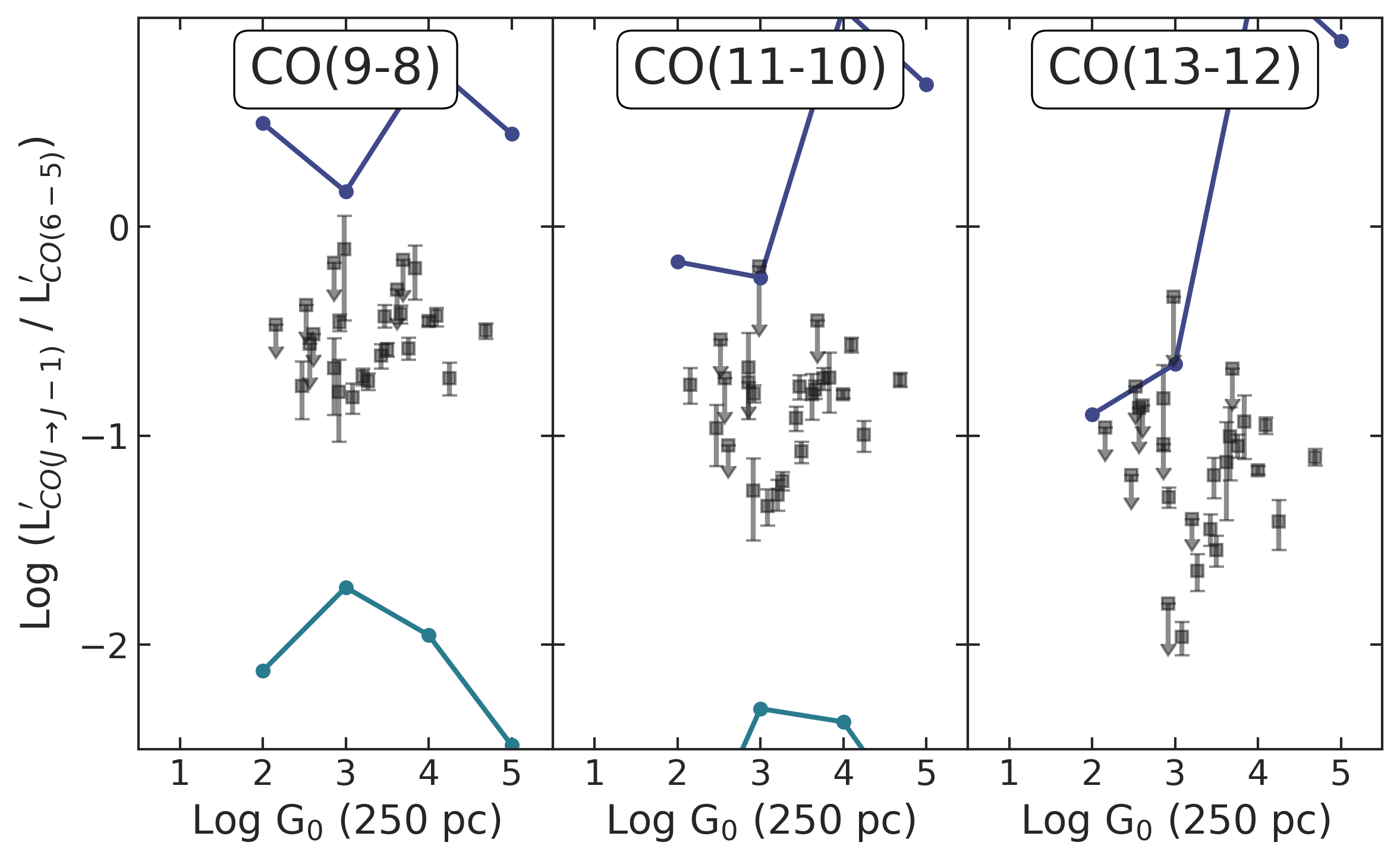}
    \end{subfigure}
\vfill
    \begin{subfigure}[t]{0.49\textwidth}
    \centering
    \includegraphics[width=\textwidth]
    {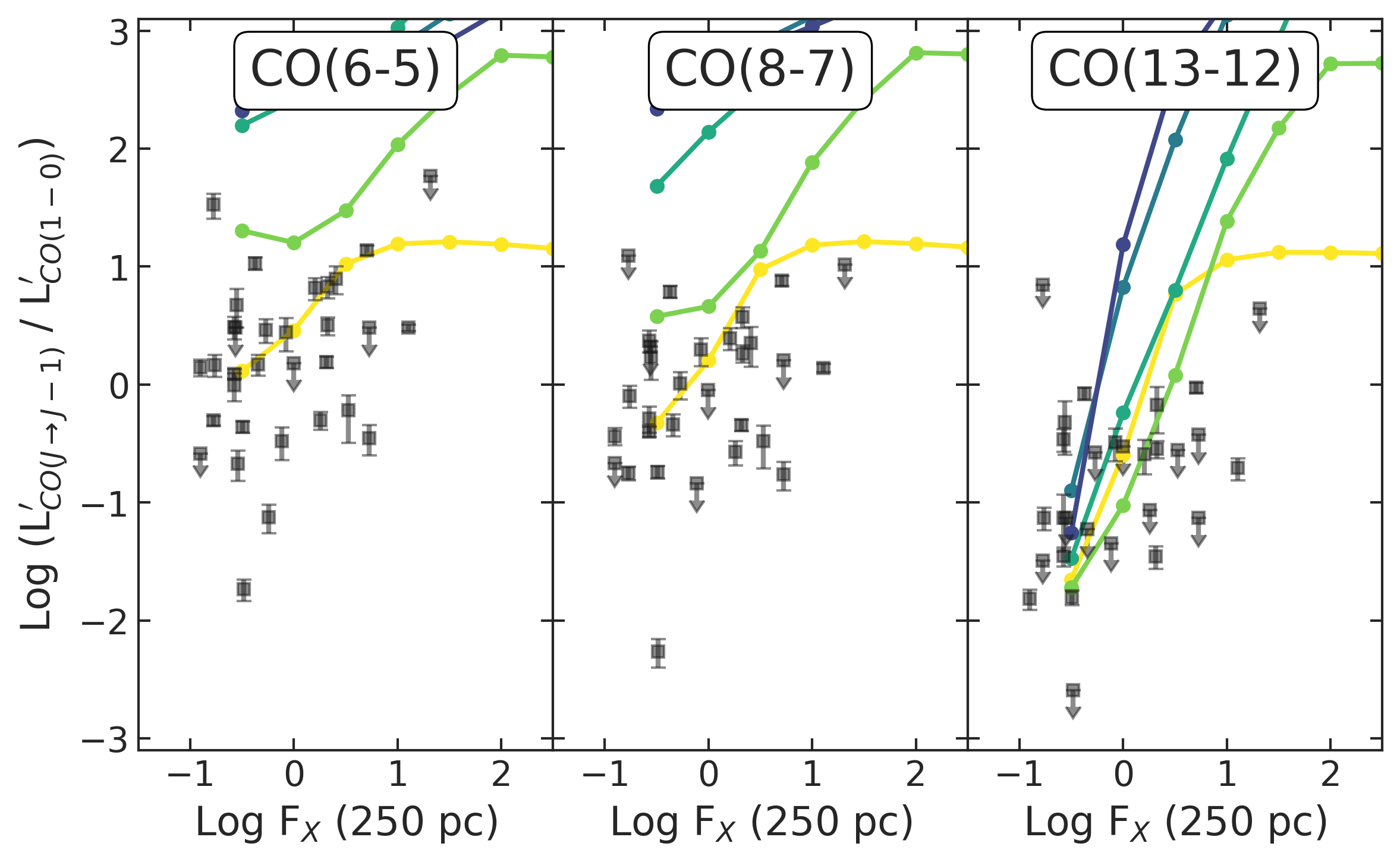}
    \caption{}
    \end{subfigure}
\hfill
    \begin{subfigure}[t]{0.49\textwidth}
    \centering
    \includegraphics[width=\textwidth]
    {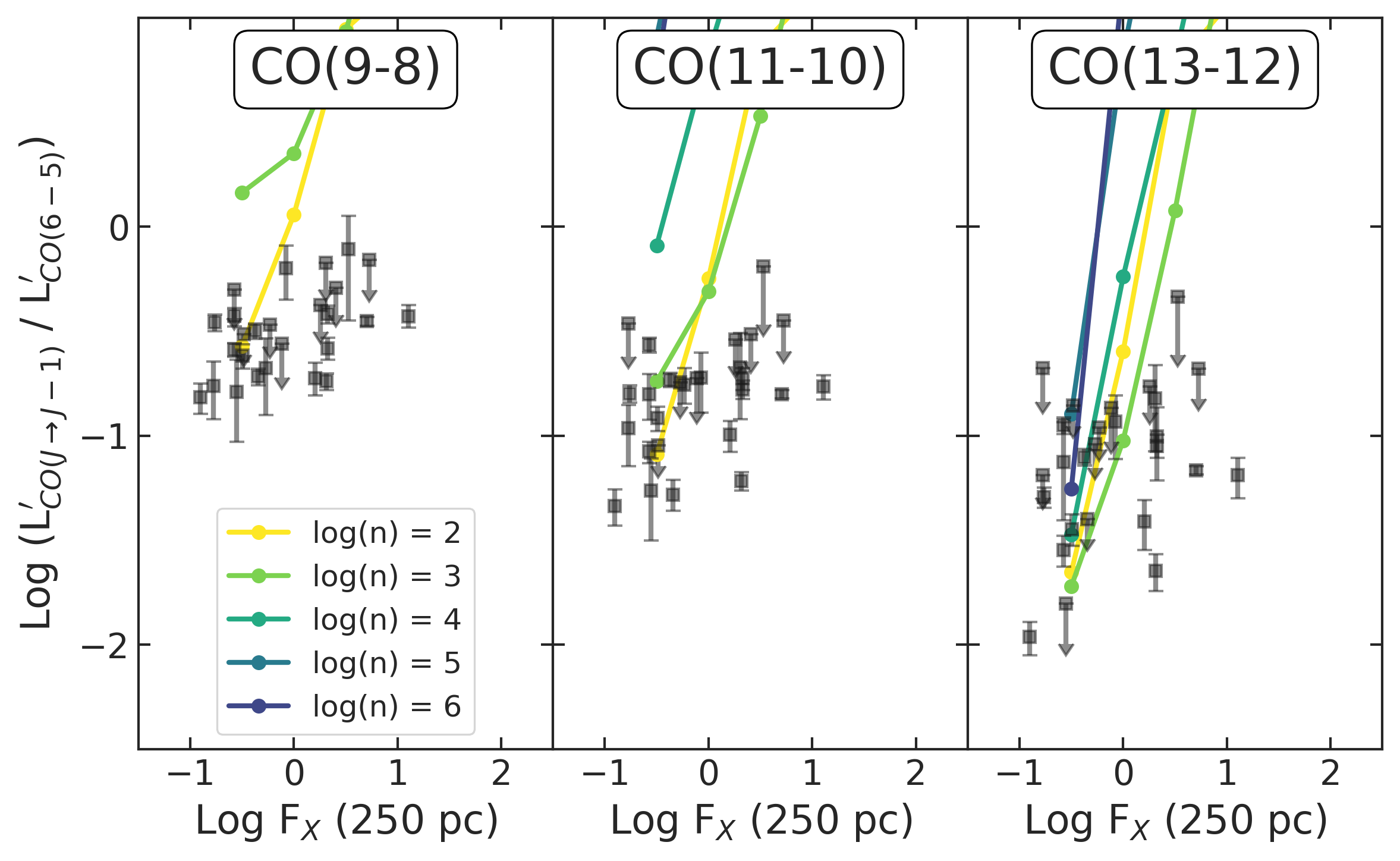}
    \end{subfigure}
\caption{
    \textit{Top-left}:
    $G_0$ vs. CO ratio to the
    nuclear ($r=250$ pc)
    fraction of CO(1--0).
    \textit{Top-right}:
    $G_0$ vs. CO ratio to the CO(6--5) line.
    \textit{Bottom-left}:
    $F_{\rm X}$ vs. CO ratio to the
    nuclear ($r=250$ pc)
    fraction of CO(1--0).
    \textit{Bottom-right}:
    $F_{\rm X}$ vs. CO ratio to the CO(6--5) line.
    In all the plots,
    the points are the same of
    Figure~\ref{fig:G0COratios} and 
    Figure~\ref{fig:FXCOratios}.
    Both $G_0$ and $F_X$ are measured
    at $r = 250$ pc.
    The coloured overplotted lines are
    from numerical \textsc{Cloudy} models 
    of different gas 
    densities $n$, namely $10^2$
    (yellow), $10^3$ (light green),
    $10^4$ (aqua green), $10^5$ (light blue)
    and $10^6$ (dark blue) cm$^{-3}$.}
\label{fig:cloudy}
\end{figure*}

Compared to the PDR results shown in
Figure~\ref{fig:G0COratios}, for the XDR
we find lower regression slopes:
$0.1 - 0.5$ for the CO(1--0) ratios,
$0 - 0.2$ for the CO(6--5) ratios.
We interpret this as a sign that neither $F_{\text{X}}$
is the dominant driver of these CO lines.
Given the physics of high-$J$ CO line emission,
which originate from warm molecular gas,
the X-ray influence was expected to show up 
in the correlation with the line ratios, 
especially those with respect to the low-$J$ CO lines,
as found by many theoretical
\citep{maloney96, meijerink05, meijerink07}
and observational 
\citep{vanderwerf10, pozzi17, mingozzi18}
works on XDR.
A plausible explanation is that 
at $r=250$ pc we are still outside
of the actual AGN sphere of influence
of the molecular gas:
several studies on Seyfert galaxies
\citep{davies07, kawakatu08, esquej14, motter21}
indeed place it within the central $r=100$ pc. 
At larger radii, we cannot isolate the contribution
of X-rays due to dilution with stellar FUV photons.
Unfortunately, our \textit{Herschel}
CO observations
have limited spatial resolution to reach
such a nuclear region,
and ALMA is still limited
to the low/mid-$J$ lines,
at least in the local Universe.

\subsection{Comparison with models}
\label{sec:models}

We use predictions from numerical models presented in 
\citet{vallini19}
to interpret the
observations, in order to
shed light on the dominant
heating source in the molecular ISM 
of our galaxies.
For this purpose, we use \textsc{Cloudy}
\citep{ferland17} to
compute the CO line intensities emerging from a 
1-D gas slab of density $n$,
illuminated by either FUV flux 
$G_0$ (PDR models) or a X-ray flux 
$F_{\rm X}$ (XDR models).
The results of these simulations
mainly apply for a single cloud,
while we are dealing with entire galaxies
(or at least their inner regions);
it is therefore especially convenient to study
the effect on the line ratios,
rather than line fluxes or luminosities,
assuming that both numerators and denominators
originate from the same area.

The gas density $n$ is a fundamental missing quantity
in our analysis of PDR and XDR.
We do have some indications
of its possible value:
from the X-ray-derived column density, we estimated
mean volume densities between
$n \approx 10^{1-3}$ cm$^{-3}$
(Section~\ref{sec:xsample})
within $r=250$ pc.
It is however possible, 
from the comparison of observed CO ratios
with PDR and XDR \textsc{Cloudy} models outputs,
to estimate the density of the dissociation region
from which the observed CO lines originate.

In the four panels of Figure~\ref{fig:cloudy},
we examine the PDR and XDR predictions,
respectively made with $\log G_0 = [2, 3, 4, 5]$ and
$\log (F_{\text{X}} / (\text{erg s}^{-1} \text{cm}^{-2}))
= [-0.5, 0, 0.5, 1, 1.5, 2, 2.5]$,
with modelled gas density
$\log (n / \text{cm}^{-3}) = [2, 3, 4, 5]$.
Again we explore the CO line ratio
to CO(1--0) and CO(6--5),
using the same three mid-/high-$J$
lines as in Figure~\ref{fig:G0COratios}
and \ref{fig:FXCOratios}.
The same plots with all the CO lines
can be found at the end of
Appendix~\ref{sec:COratiofull}.
The modelled points are plotted in the panels
of Figure~\ref{fig:cloudy}, colour coded with $n$.

In the PDR case, almost all 
our galaxies are reproduced
considering densities in the $n=10^{5-6}$ cm$^{-3}$ range,
except for the line ratios
up to CO(6--5), as can be seen
on the leftmost panel
($L^\prime_{CO(6-5)} / L^\prime_{CO(1-0)}$)
of Figure~\ref{fig:cloudy},
and even better in the first lines of
Figure~\ref{fig:PDRCO1full}~and~\ref{fig:PDRCO6full}.
Previous PDR studies
did not find such high densities.
The only exception is Mrk 231,
for which \cite{vanderwerf10} obtained
a warm PDR component with $G_0 = 10^{3.5}$
and $n=10^5$ cm$^{-3}$;
however, such a high density is necessary
to reproduce the mid-$J$ emission,
while a colder PDR component, with $n=10^{3.5}$ cm$^{-3}$,
reproduces the low-$J$ emission and accounts
for most of the gas volume.
\cite{diazsantos17} observed instead
that on average, and on the 
scale of the whole galaxy,
local (U)LIRGs start from a minimum
$G_0 / n \sim 10^{-1}$, and that
this ratio increases with
the IR luminosity surface density;
this would place an upper limit
to the gas density at a fixed $G_0$.
In the top panels of Figure~\ref{fig:cloudy}, instead,
our galaxies, for $J_{\text{upp}} \geq 8$,
lie in the range $\log(G_0 / n) = [-4, -1]$,
given the modelled gas densities.
It is necessary for PDR models to have high densities
to produce bright mid-$J$ transitions \citep{vallini18},
and it is known \citep[e.g.][]{mckee07} 
that such densities are typical of clumps and cores 
in single star-forming molecular clouds
\citep[as shown by][in e.g. the Orion Bar]{joblin18}.
Nonetheless, it is 
unlikely that the central 500 pc
of galaxies have an average gas density
of $10^{5-6}$ cm$^{-3}$, so we expect
these high-density regions 
to have a very low volume filling factor.

In the XDR case, on the contrary,
the models with low density
($n \approx 10^{2-3}$ cm$^{-3}$)
can reproduce the observed CO line ratios,
at least in the regions of the parameters space
where the lines with such densities
are clearly separable from the others.
This result is in line with the densities
($n \approx 10^{1-3}$ cm$^{-3}$)
calculated from the X-ray-derived $N_{\rm H}$,
and from what we expect from
the available XDR studies
for local (U)LIRGs
\citep{vanderwerf10, pozzi17, mingozzi18}.
From Figure~\ref{fig:cloudy}
it is clear that the observed 
high-$J$ line ratios
(especially $J_{\text{upp}} \geq 12$)
can be reproduced by either a high $F_X$
or a high $n$,
a degeneracy also found in
the semi-analytic model
by \cite{vallini19}.
However, both our high-$J$ line ratios
and our calculated $F_{\text{X}}$
are lowered by the nuclear radius
we are using ($r = 250$ pc),
so a detailed numerical modelling
at different distances from the AGN
is needed to really see the impact
of XDR on the molecular emission.

We note here that stars and AGN
can also affect the heating of molecular gas through
outflows/winds, resulting in shock-heated regions
\citep{kazandjian12, aalto12, burillo14} 
where the brightness of high-J CO lines is enhanced too.
Disentangling the contribution of shock heating 
from that produced in XDRs is a challenging task 
\citep{hollenbach89, meijerink13, mingozzi18}.
However, the study of mechanical heating is beyond 
the scope of this paper.

\section{Conclusions}
In this paper,
we investigate the relative
impact of star formation and AGN activity
on the CO rotational line emission.
In this respect, we collect multiwavelength
(mm, IR and X-ray) data for a sample
of 35 local active galaxies.
The sources are selected
with a well-sampled CO SLED
(from $J=1-0$ to $J=13-12$)
and intrinsic $L_X \geq 10^{42}$
erg s$^{-1}$ in the 2--10 keV range.
From the multiband data we derive,
in a homogeneous way, key integrated physical quantities, 
as the molecular gas mass ($M_{\text{mol}}$),
the star formation FUV flux ($G_0$)
and the AGN X-ray flux, $F_{\rm X}$. 
Moreover, by analysing the ALMA images
of the highest available CO emission,
we estimate the emitting area of mid-$J$/high-$J$
CO lines,
finding it concentrated within
$r = 250$ pc from the center.
To determine whether AGN activity
influences the molecular gas
in its vicinity,
we measure FUV and X-ray radiation,
producing PDR and XDR, respectively,
from the observational
data in a self-consistent way.
The FUV flux is parametrized in terms of $G_0$,
gauged from the 70 $\mu$m,
spatially resolved, dust emission,
the $F_X$ is calculated
from the intrinsic $L_X$. 
Our main results can be summarized
as follows:

\renewcommand{\labelenumi}{\arabic{enumi}.}
\begin{enumerate}
\item On the kpc-scale of the whole galaxy
(namely within a median
$r_{\text{CO}} = 3.1^{+2.1}_{-1.5}$ kpc)
we do not find measurable evidence for the
AGN influence on the star formation. 
Our sample results well mixed with other samples
of non-active galaxies on the Schmidt-Kennicutt
($\Sigma_{\text{mol}}$ vs. $\Sigma_{\text{SFR}}$) plane.
If we use a Milky Way CO-to-H$_2$ conversion factor
$\alpha_{\text{CO}} = 4.3$
M$_{\odot}$ (K km s$^{-1}$ pc$^{-2}$)$^{-1}$,
we find a median
$\log (M_{\text{mol}} / M_{\odot}) = 9.9^{+0.3}_{-0.8}$
for our sample, and a median depletion time
$\log (\tau_{\text{depl}} / \text{yr}) = 8.9^{+0.4}_{-0.6}$.
\item We measure within $r=250$ pc
the irradiation of PDR and XDR by deriving $G_0$ and $F_X$,
finding $\log G_0 = 3.1^{+0.7}_{-0.8}$ and
$\log (F_{\text{X}} / (\text{erg s}^{-1} \text{cm}^{-2}))
= -0.1_{-0.5}^{+0.8}$ for our sample.
These values are comparable with
the literature for local active galaxies,
for both observational and theoretical works.
\item We find weak correlations
between $G_0$, $F_{\text{X}}$
and two different CO line ratios,
namely to the nuclear ($r=250$ pc) fraction of
CO(1--0) and to CO(6--5).
Therefore, neither $G_0$ nor $F_{\text{X}}$ alone
can produce the observed molecular emission.
\item From the comparison of CO emission
and observed $G_0$
with grids of PDR numerical models,
we can conclude that PDR emission
can reproduce
observed high-$J$ line ratios
only assuming unlikely extreme gas densities
($n > 10^5$ cm$^{-3}$),
while it is more efficient at moderate
densities ($n \sim 10^{3-4}$ cm$^{-3}$)
up to CO(6--5).
\item From the comparison 
between XDR observations
and models, we find that
$F_{\text{X}}$ can reproduce the
observed low-/mid-$J$ CO line ratios
only at low densities 
($n \sim 10^{2}$ cm$^{-3}$),
similar to those estimated
from X-ray column densities
($n \sim 10^{1-3}$ cm$^{-3}$).
At high-$J$ we find increasing
(with $J$) degeneracy between
$F_{\text{X}}$ and $n$,
so we can not find a typical
gas density for our sample.
This is probably an indication
that the nuclear scale at which 
we are considering the XDR
is still too large to see
a strong AGN effect on the CO SLED.
\end{enumerate}

From our analysis, we conclude that,
on scales of $\approx$250~pc from the galaxy center,
a mix of PDR and XDR is necessary to explain the
observed CO emission,
since neither of them is the dominant mechanism.
The use of the CO SLED to disentangle the contribution of FUV and/or X-rays photons to the molecular gas heating in local galaxies is currently limited by the
low spatial resolution
at the high-$J$ frequencies
($\sim 17$ arcsec for CO(13--12) with
\textit{Herschel}/PACS).
Conversely, high-$z$ galaxies have their high-$J$ CO emission
redshifted into the observation
bands of ALMA and NOEMA, which are able to reach
sub-arcsec resolution.
These extreme CO lines
have been observed and modelled
already by several works
\citep{gallerani14, carniani19, pensabene21}.
It would be therefore interesting
to extend the analysis performed in this paper 
on a high-redshift sample
of active galaxies with spatially resolved CO emission, and assess possible differences with local AGN.

\section*{Acknowledgements}

We thank the anonymous referee for the helpful comments
that increased the quality of this paper.
We acknowledge use of
APLpy \citep{aplpy1, aplpy2},
Astropy \citep{astropy1, astropy2},
Matplotlib \citep{matplotlib},
NumPy \citep{numpy},
Pandas \citep{pandas},
Photutils \citep{photutils},
Python \citep{python3},
Seaborn \citep{seaborn},
Scikit-learn \citep{sklearn},
SciPy \citep{scipy}.
We acknowledge the usage of the HyperLeda database (http://leda.univ-lyon1.fr), \cite{hyperleda}.
Optical images of galaxies are based on observations made with the NASA/ESA Hubble Space Telescope, and obtained from the Hubble Legacy Archive, which is a collaboration between the Space Telescope Science Institute (STScI/NASA), the Space Telescope European Coordinating Facility (ST-ECF/ESA) and the Canadian Astronomy Data Centre (CADC/NRC/CSA).
This research has made use of the services of the ESO Science Archive Facility.
We acknowledge the use of DSS (Digitized Sky Survey) images.
The DSS was produced at the Space Telescope Science Institute under U.S. Government grant NAG W-2166. The images of these surveys are based on photographic data obtained using the Oschin Schmidt Telescope on Palomar Mountain and the UK Schmidt Telescope.
This research has made use of Aladin Sky Atlas developed at CDS, Strasbourg Observatory, France, 
\cite{aladin1} and \cite{aladin2}.
FE and FP acknowledge support from grant PRIN MIUR 2017- 20173ML3WW$\_$001.
We acknowledge support from the INAF mainstream 2018 program 
``Gas-DustPedia: A definitive view of the ISM in the Local Universe".

\section*{Data Availability}

The data underlying this article 
were accessed from the ALMA Archive
(\url{https://almascience.eso.org/asax/}),
from the JVO portal (\url{http://jvo.nao.ac.jp/portal})
operated by the NAOJ,
and the NASA/IPAC Infrared Science Archive
(specifically
\url{https://irsa.ipac.caltech.edu/data/Herschel/HHLI/overview.html}), 
which is funded by the National Aeronautics and Space 
Administration and operated by the California Institute of 
Technology
The derived data generated in this research
will be shared on reasonable request 
to the corresponding author.


\bibliographystyle{mnras}
\bibliography{AGNimpactCO}



\appendix

\section{CO line ratios}
\label{sec:COratiofull}

In this section we show the CO luminosity
ratios, both with denominators
the CO(1--0) and the CO(6--5) luminosity.
The CO(1--0) luminosities have been corrected
to take into account only the emission
up to $r=250$ pc from the center 
of the galaxies
(with Equation~\ref{eq:COprofile}).
Firstly we plot the luminosity ratios
against the FUV flux $G_0$ and
the X-ray flux $F_{\text{X}}$,
fitting the points with a
regression line, respectively
as in Figures~\ref{fig:G0COratios}
and \ref{fig:FXCOratios}.
The details can be found in 
Sections~\ref{sec:pdr}
and \ref{sec:xdr}.
Secondly, we plot
the same points but with
the \texttt{Cloudy} models
at different gas densities superimposed,
as in Figure~\ref{fig:cloudy},
and as explained in detail
in Section~\ref{sec:models}.

\begin{figure*}
	\includegraphics[width=\textwidth]{
	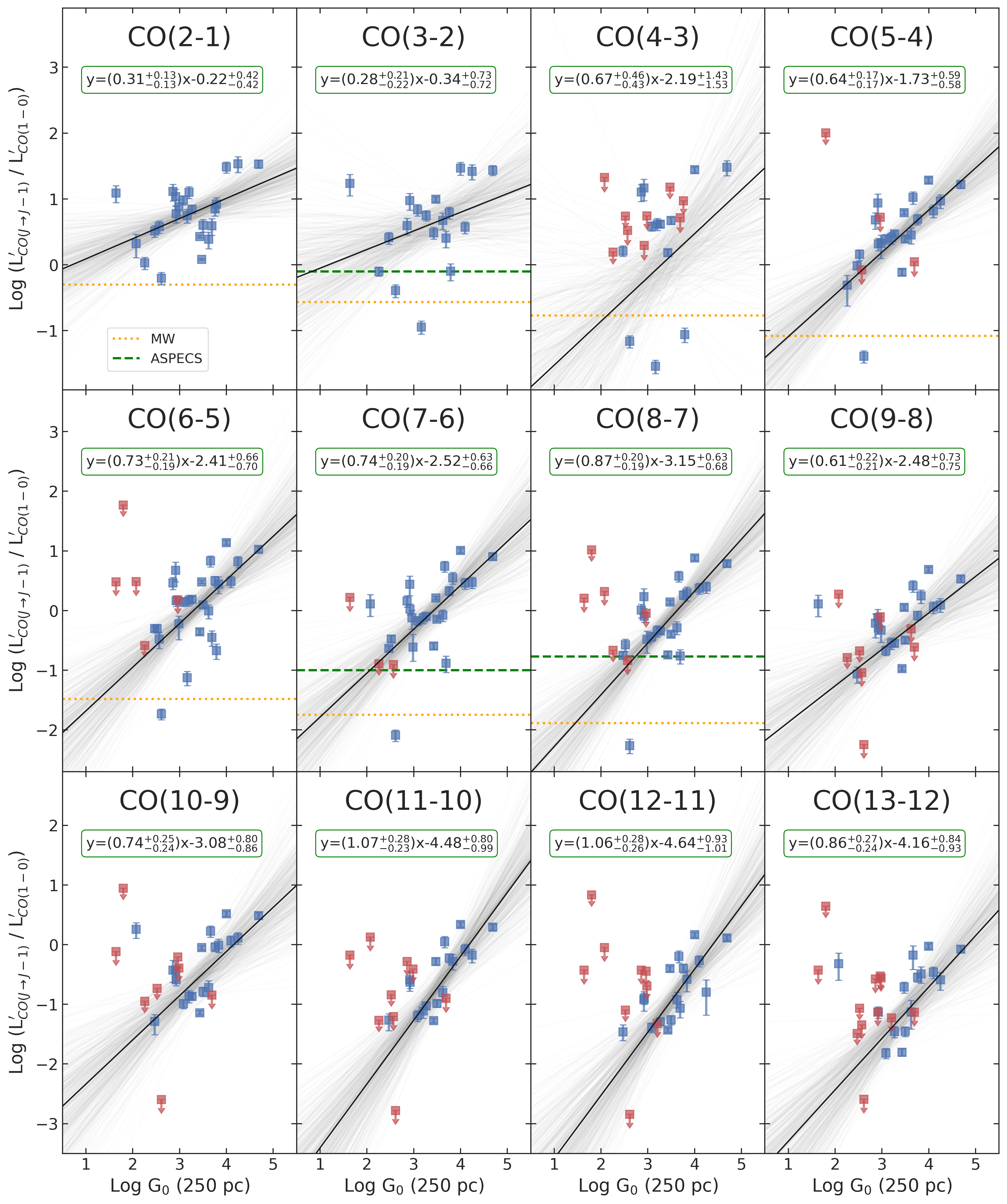}
    \caption{CO line ratios, with respect
    to the CO(1--0) line, vs. $G_0$.
    The $x$-axis is the Habing field
    $G_0$ (for $r=250$ pc).
    The $y$-axis is the luminosity ratio
    $L^{\prime}_{\text{CO}(J \rightarrow J-1)} / L^{\prime}_{\text{CO}(1 \rightarrow 0)}$
    to the nuclear ($r=250$ pc)
    fraction of CO(1--0).
    The luminosities $L^{\prime}$ are in units of
    K km s$^{-1}$ pc$^{-2}$, and
    $J$ is indicated on the top of each panel.
    Blue squares indicate $3 \sigma$ detections
    in both lines;
    red squares with downward arrow
    indicate less than $3 \sigma$ in the
    higher-$J$ line (i.e. censored data).
    The lines are
    regression fits to the observed data:
    solid black line is the median Linmix regression,
    thin shaded green lines
    show fits drawn from the posterior distribution
    of Linmix regression.
    When available, the Milky Way
    (dotted orange line, data from \citealt{fixsen99})
    and the ASPECS AGN
    (green dashed line, data from \citealt{boogaard20})
    CO ratios are also shown.
    }
    \label{fig:G0CO1full}
\end{figure*}

\begin{figure*}
	\includegraphics[width=\textwidth]{
	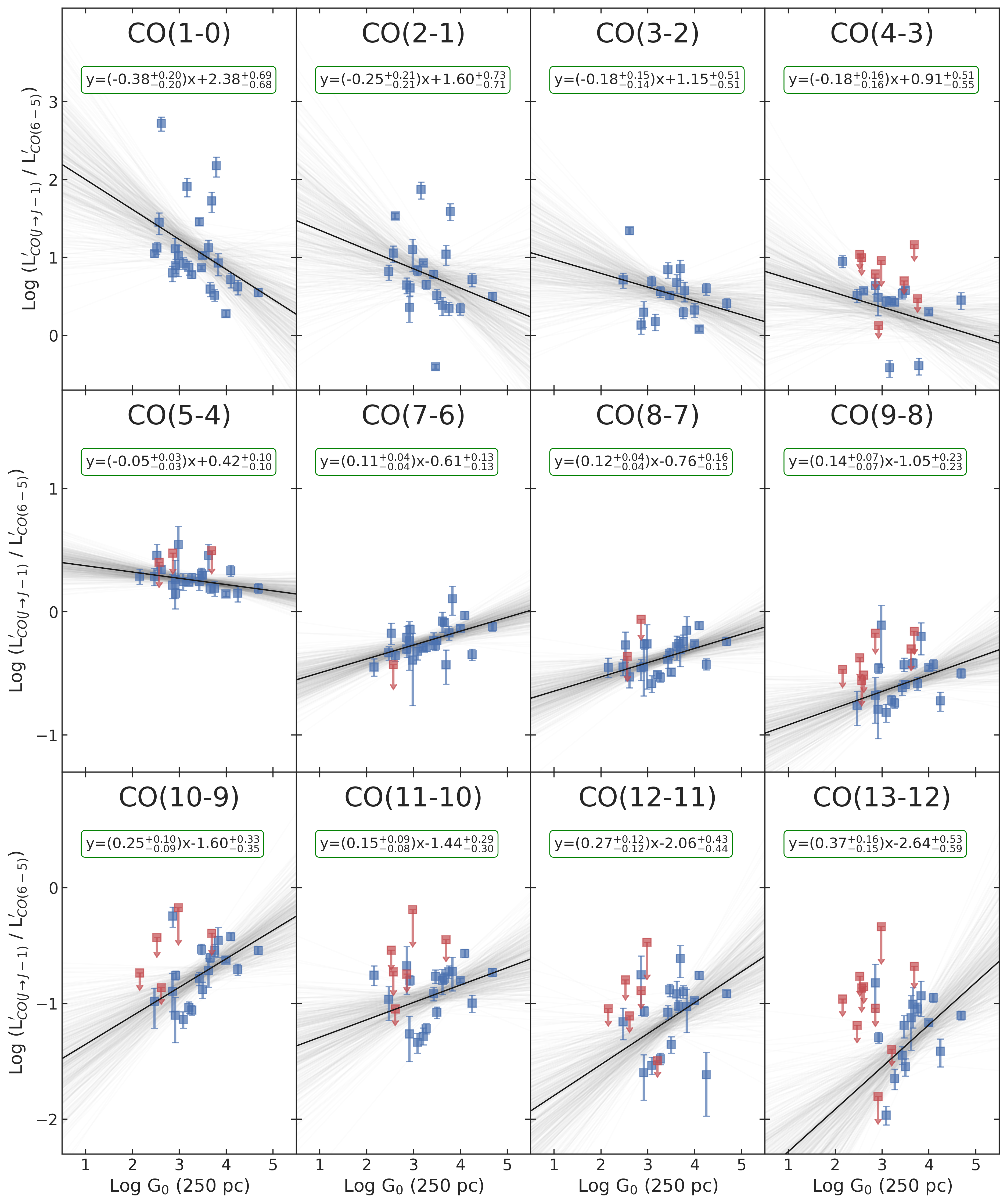}
    \caption{CO line ratios, with respect
    to the CO(6--5) line, vs. $G_0$.
    The $x$-axis is the Habing field
    $G_0$ (for $r=250$ pc).
    The $y$-axis is the luminosity ratio
    $L^{\prime}_{\text{CO}(J \rightarrow J-1)} / L^{\prime}_{\text{CO}(6 \rightarrow 5)}$
    to the CO(6--5) line.
    Data points and lines are
    described in Figure~\ref{fig:G0CO1full}.
    }
    \label{fig:G0CO6full}
\end{figure*}

\begin{figure*}
	\includegraphics[width=\textwidth]{
	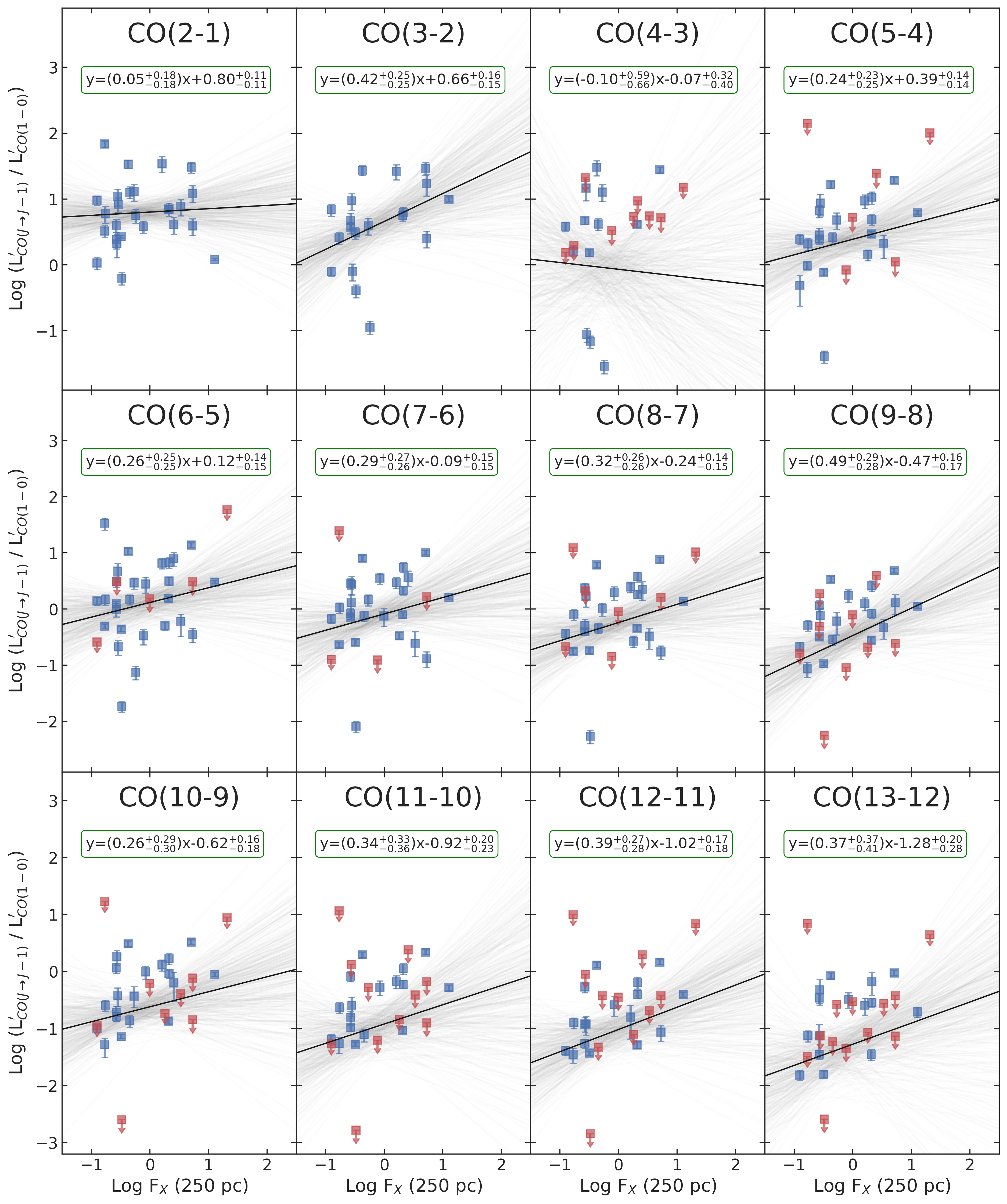}
    \caption{CO line ratios, with respect
    to the CO(1--0) line, vs. $F_X$.
    The $x$-axis is $F_X$ (for $r=250$ pc),
    in units of erg s$^{-1}$ cm$^{-2}$.
    The $y$-axis is the luminosity ratio
    $L^{\prime}_{\text{CO}(J \rightarrow J-1)} / L^{\prime}_{\text{CO}(1 \rightarrow 0)}$
    to the nuclear ($r=250$ pc)
    fraction of CO(1--0).
    Data points and lines are
    described in Figure~\ref{fig:G0CO1full}.
    }
    \label{fig:FXCO1full}
\end{figure*}

\begin{figure*}
	\includegraphics[width=\textwidth]{
	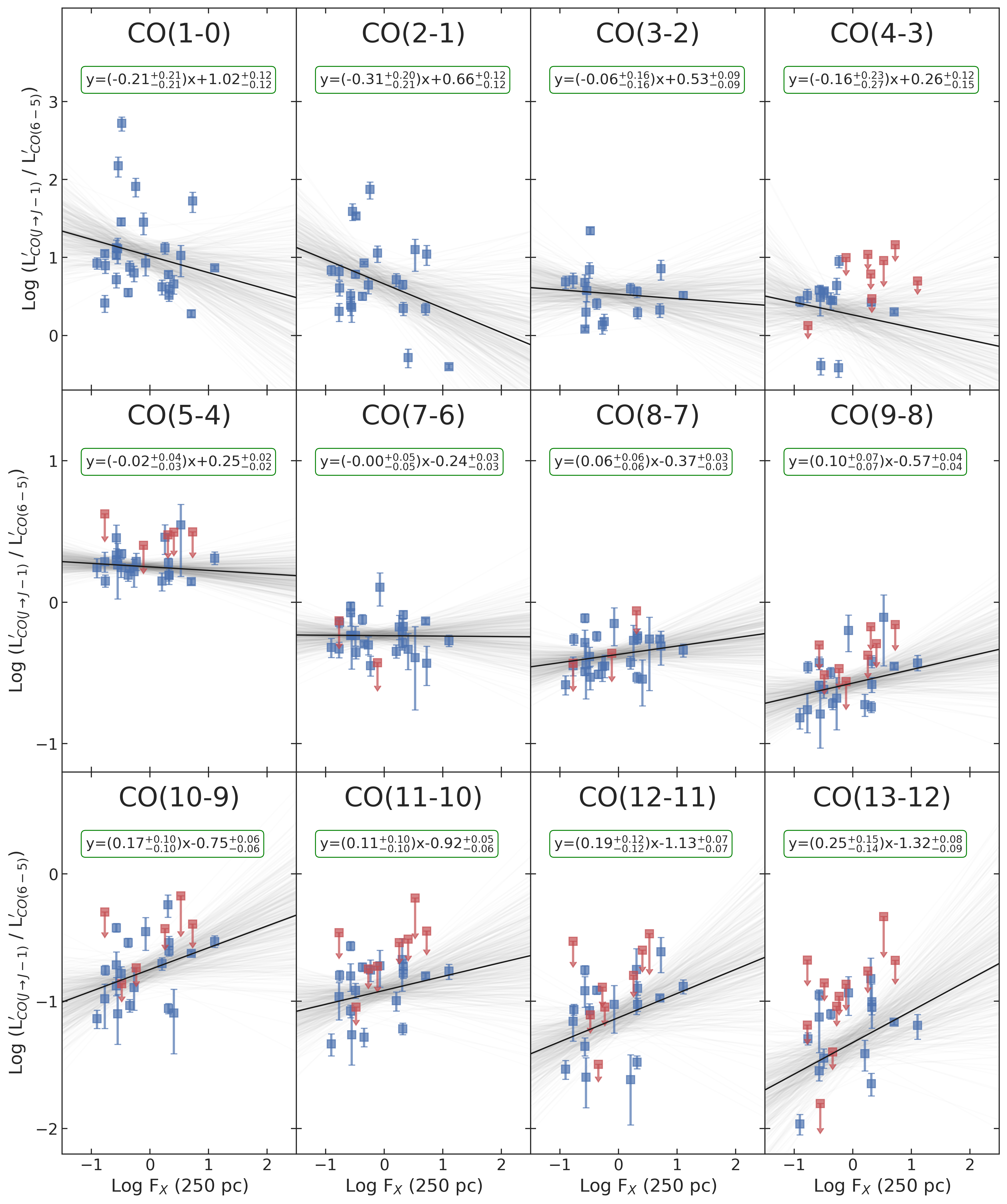}
    \caption{CO line ratios, with respect
    to the CO(6--5) line, vs. $F_X$.
    The $x$-axis is $F_X$ (for $r=250$ pc),
    in units of erg s$^{-1}$ cm$^{-2}$.
    The $y$-axis is the luminosity ratio
    $L^{\prime}_{\text{CO}(J \rightarrow J-1)} / L^{\prime}_{\text{CO}(6 \rightarrow 5)}$
    to the CO(6--5) line.
    Data points and lines are
    described in Figure~\ref{fig:G0CO1full}.
    }
    \label{fig:FXCO6full}
\end{figure*}

\begin{figure*}
	\includegraphics[width=\textwidth]{
	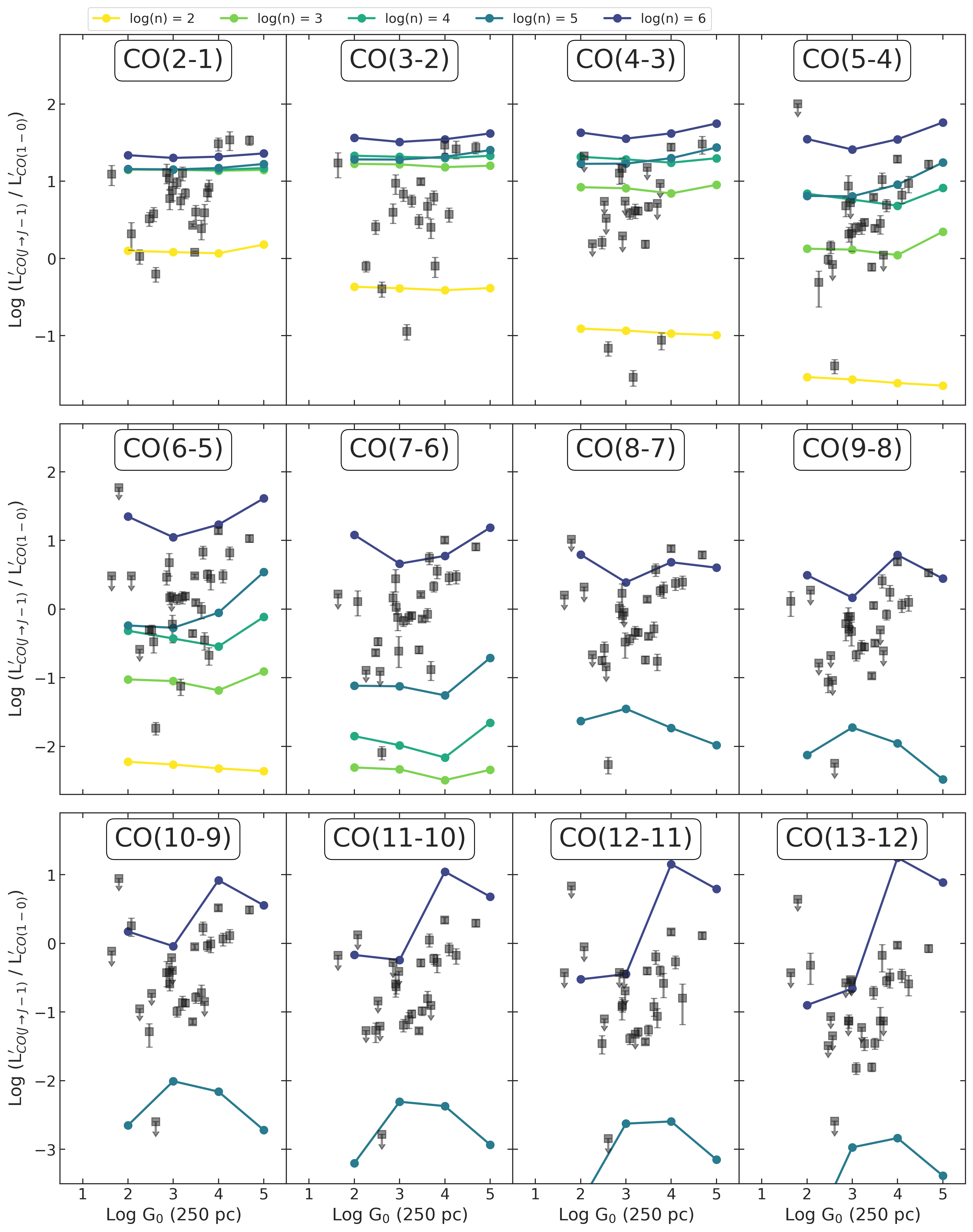}
    \caption{CO line ratios, with respect
    to the CO(1--0) line, vs. $G_0$.
    The $x$-axis is the Habing field
    $G_0$ (for $r=250$ pc).
    The $y$-axis is the luminosity ratio
    $L^{\prime}_{\text{CO}(J \rightarrow J-1)} / L^{\prime}_{\text{CO}(1 \rightarrow 0)}$
    to the nuclear ($r=250$ pc)
    fraction of CO(1--0).
    The luminosities $L^{\prime}$ are in units of
    K km s$^{-1}$ pc$^{-2}$, and
    $J$ is indicated on the top of each panel.
    Squares with downward arrow
    indicate less than $3 \sigma$ 
    detections in the
    higher-$J$ line (i.e. censored data).
    The colored overplotted lines are \textsc{Cloudy}
    numerical models at different gas 
    densities, namely $10^2$
    (yellow), $10^3$ (light green),
    $10^4$ (dark green), $10^5$ (blue)
    and $10^6$ (purple) cm$^{-3}$.
    }
    \label{fig:PDRCO1full}
\end{figure*}

\begin{figure*}
	\includegraphics[width=\textwidth]{
	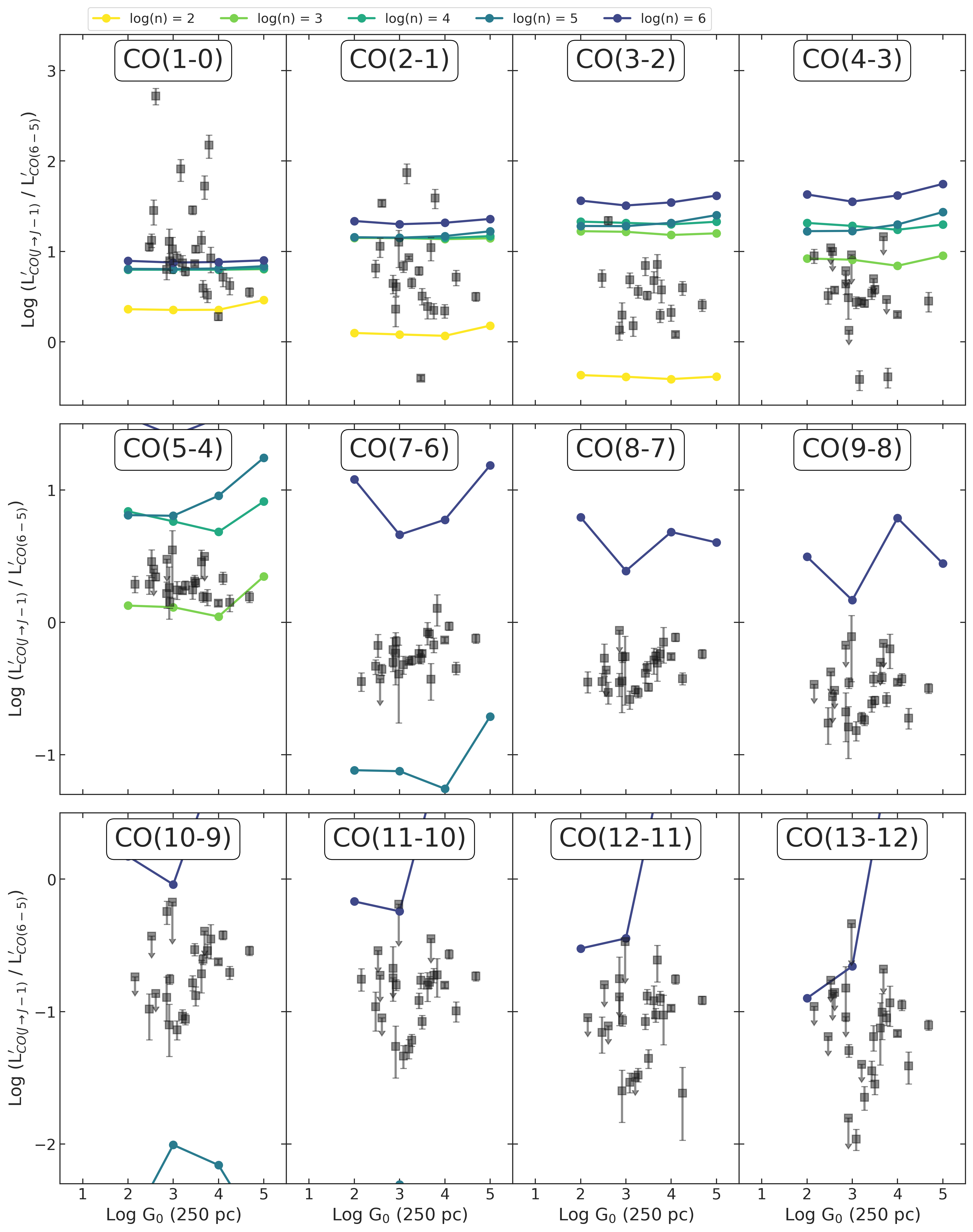}
    \caption{CO line ratios, with respect
    to the CO(6--5) line, vs. $G_0$.
    The $x$-axis is the Habing field
    $G_0$ (for $r=250$ pc).
    The $y$-axis is the luminosity ratio
    $L^{\prime}_{\text{CO}(J \rightarrow J-1)} / L^{\prime}_{\text{CO}(6 \rightarrow 5)}$
    to the CO(6--5) line.
    Data points and lines are
    described in Figure~\ref{fig:PDRCO1full}.
    }
    \label{fig:PDRCO6full}
\end{figure*}

\begin{figure*}
	\includegraphics[width=\textwidth]{
	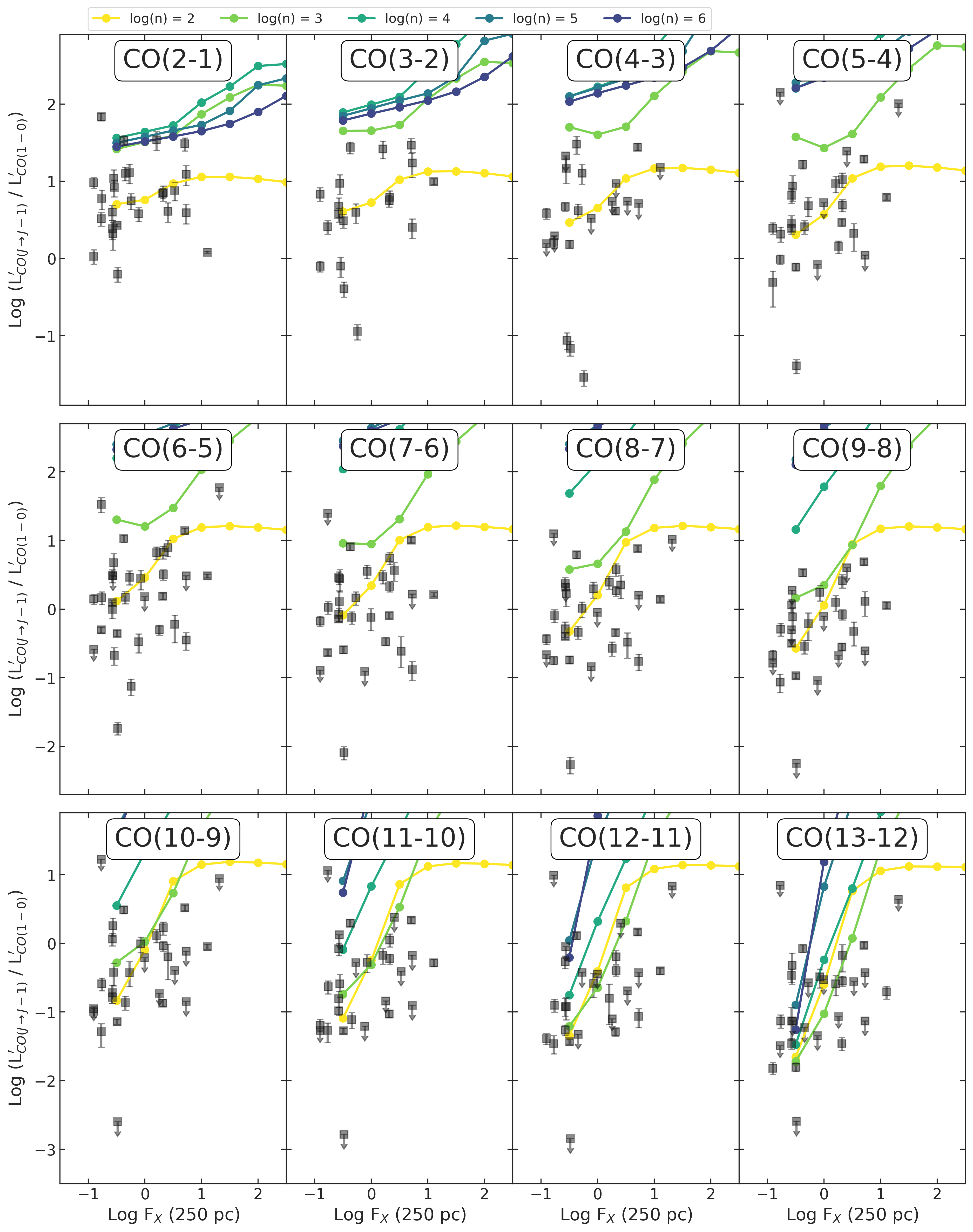}
    \caption{CO line ratios, with respect
    to the CO(1--0) line, vs. $F_X$.
    The $x$-axis is $F_X$ (for $r=250$ pc),
    in units of erg s$^{-1}$ cm$^{-2}$.
    The $y$-axis is the luminosity ratio
    $L^{\prime}_{\text{CO}(J \rightarrow J-1)} / L^{\prime}_{\text{CO}(1 \rightarrow 0)}$
    to the nuclear ($r=250$ pc)
    fraction of CO(1--0).
    Data points and lines are
    described in Figure~\ref{fig:PDRCO1full}.
    }
    \label{fig:XDRCO1full}
\end{figure*}

\begin{figure*}
	\includegraphics[width=\textwidth]{
	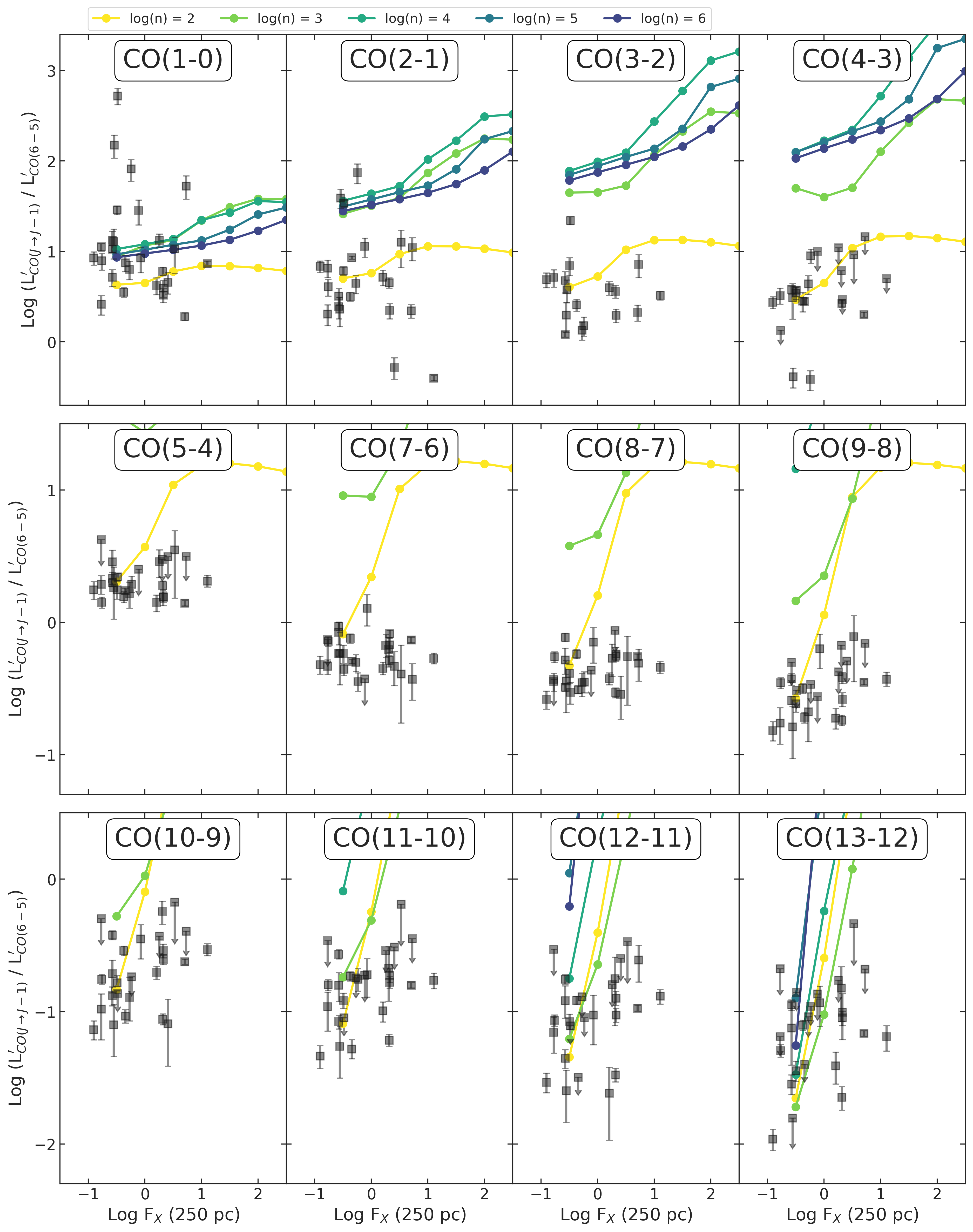}
    \caption{CO line ratios, with respect
    to the CO(6--5) line, vs. $F_X$.
    The $x$-axis is $F_X$ (for $r=250$ pc),
    in units of erg s$^{-1}$ cm$^{-2}$.
    The $y$-axis is the luminosity ratio
    $L^{\prime}_{\text{CO}(J \rightarrow J-1)} / L^{\prime}_{\text{CO}(6 \rightarrow 5)}$
    to the CO(6--5) line.
    Data points and lines are
    described in Figure~\ref{fig:PDRCO1full}.
    }
    \label{fig:XDRCO6full}
\end{figure*}

\section{CO(6--5) atlas}
\label{sec:CO6atlas}

In this section we present the rest
(in addition to Figure~\ref{fig:N34}) of the images
of CO(6--5) emission for our sample galaxies.
All the CO(6--5) data cubes are from the ALMA Archive,
already calibrated, cleaned, and when available,
primary-beam corrected.
Using CASA 5.6 \citep{mcmullin07},
we produce the moment 0 map from the data cubes
with the task \texttt{immoments}.
We then plot the ALMA CO(6--5) contours
over the optical image of the galaxy.

\begin{figure*}
    \includegraphics[width=\textwidth]{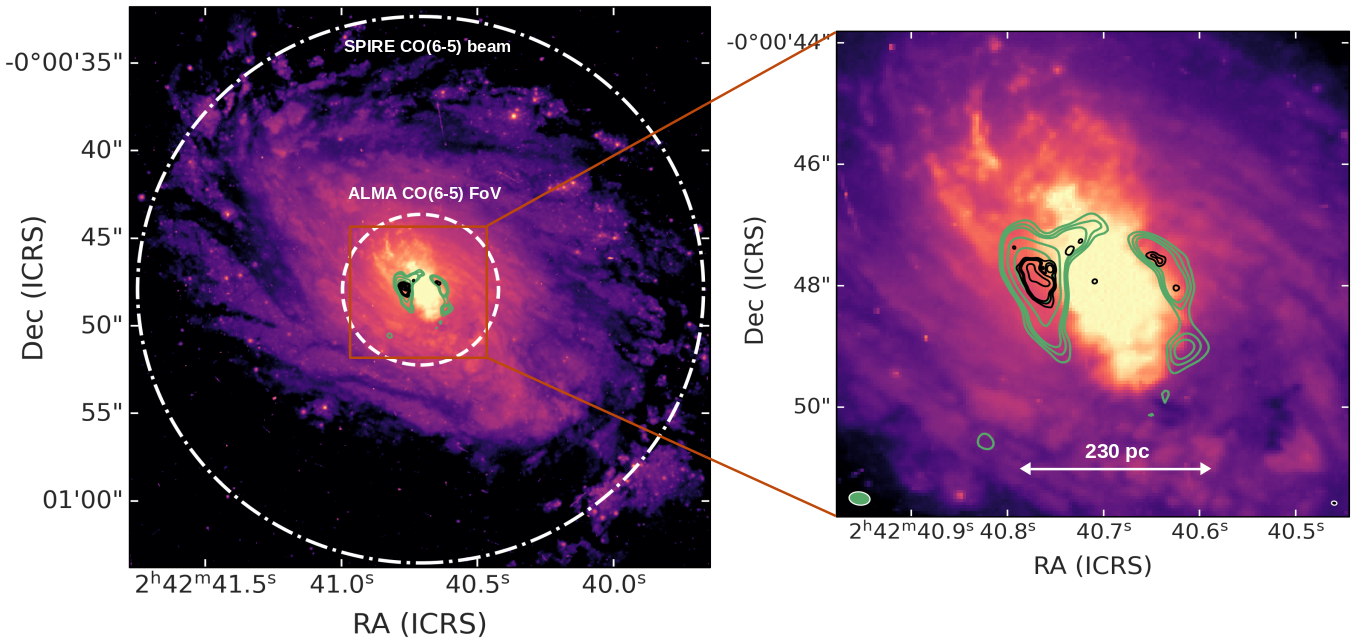}
    \caption{
    \textit{Left panel:} HST WFPC2 F606W image 
    of NGC~1068 (from \protect\citealt{malkan98})
    with superimposed
    the contours of two ALMA CO(6--5) observations,
    in green at the resolution of 250~mas
    (project 2011.0.00083.S, PI: García-Burillo), 
    in black of 90~mas
    (project 2013.1.00014.S, PI: Elitzur).
    Both the contours are at the respective
    (3, 4, 5, 10, 20) $\times \, \sigma$,
    where $\sigma = 6.2$ Jy beam$^{-1}$ km s$^{-1}$ for the green lines
    and $\sigma = 1.1$ Jy beam$^{-1}$ km s$^{-1}$ for the black lines.
    The inner white dashed circle indicates the FoV
    of both ALMA observations, with a radius of 4\farcs3 ($\sim$340~pc),
    while the outer dash-dotted circle represents
    the \textit{Herschel}/SPIRE-FTS beam FWHM
    for CO(6--5) observations,
    with a 15\farcs6 radius.
    \textit{Right panel:} zoom of the inner 670~pc.
    Restored ALMA beams of the 250 and 90~mas
    images are shown as ellipses with white edges,
    at the bottom left (with the green area) 
    and right (with the black area), respectively.
    The 250~mas ALMA image has not been primary-beam corrected.
    }
    \label{fig:N1068}
\end{figure*}

\begin{figure*}
    \includegraphics[width=\textwidth]{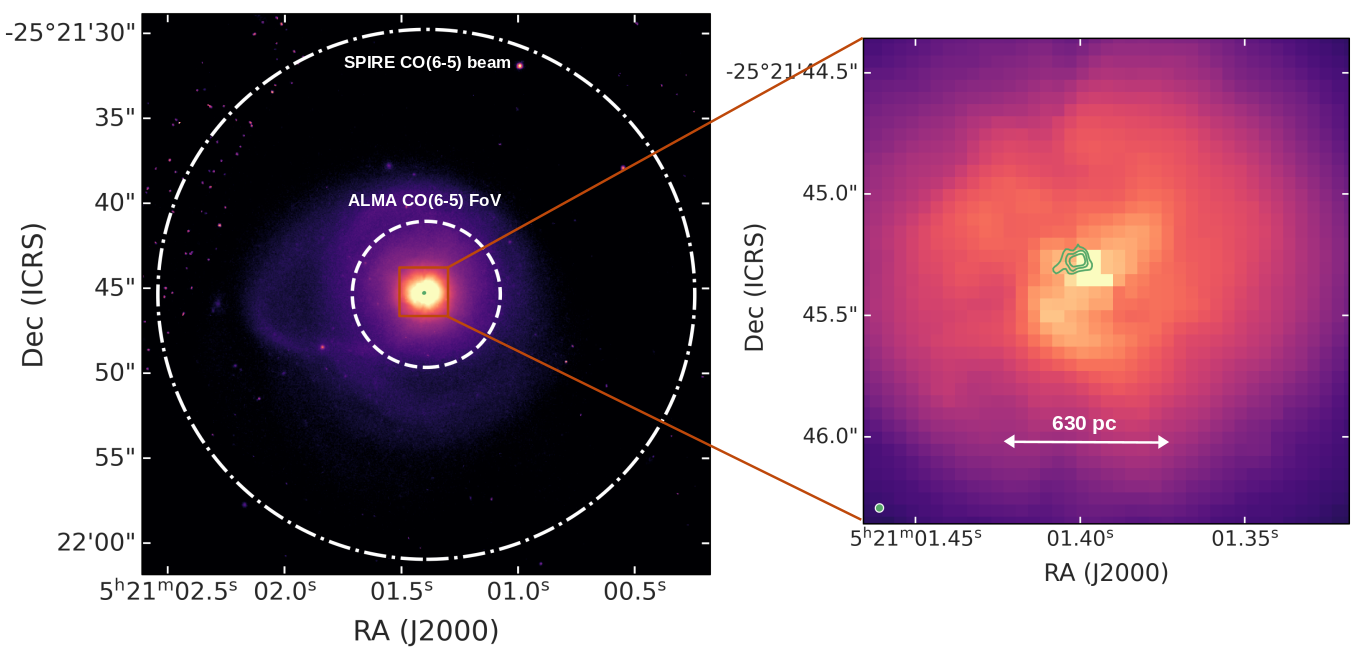}
    \caption{
    \textit{Left panel:} HST ACS F814W image 
    of IRAS~F05189--2524 (from \protect\citealt{evansPI}),
    with superimposed, in green,
    the contours of ALMA CO(6--5) moment 0
    at the resolution of 40~mas
    (project 2016.1.01223.S, PI: Baba).
    The contours are drawn at
    (3, 4, 5, 10, 20) $\times \, \sigma$,
    where $\sigma = 0.55$ Jy beam$^{-1}$ km s$^{-1}$.
    The inner white dashed circle indicates the FoV
    of both ALMA observations, with a radius of 4\farcs3 ($\sim$3.6~kpc),
    while the outer dash-dotted circle represents
    the \textit{Herschel}/SPIRE-FTS beam FWHM
    for CO(6--5) observations,
    with a 15\farcs6 radius.
    \textit{Right panel:} zoom of the inner 1.7~kpc.
    The restored ALMA beam
    is shown as a green ellipse with white edges
    at the bottom left.
    }
    \label{fig:F05189}
\end{figure*}

\begin{figure*}
    \includegraphics[width=\textwidth]{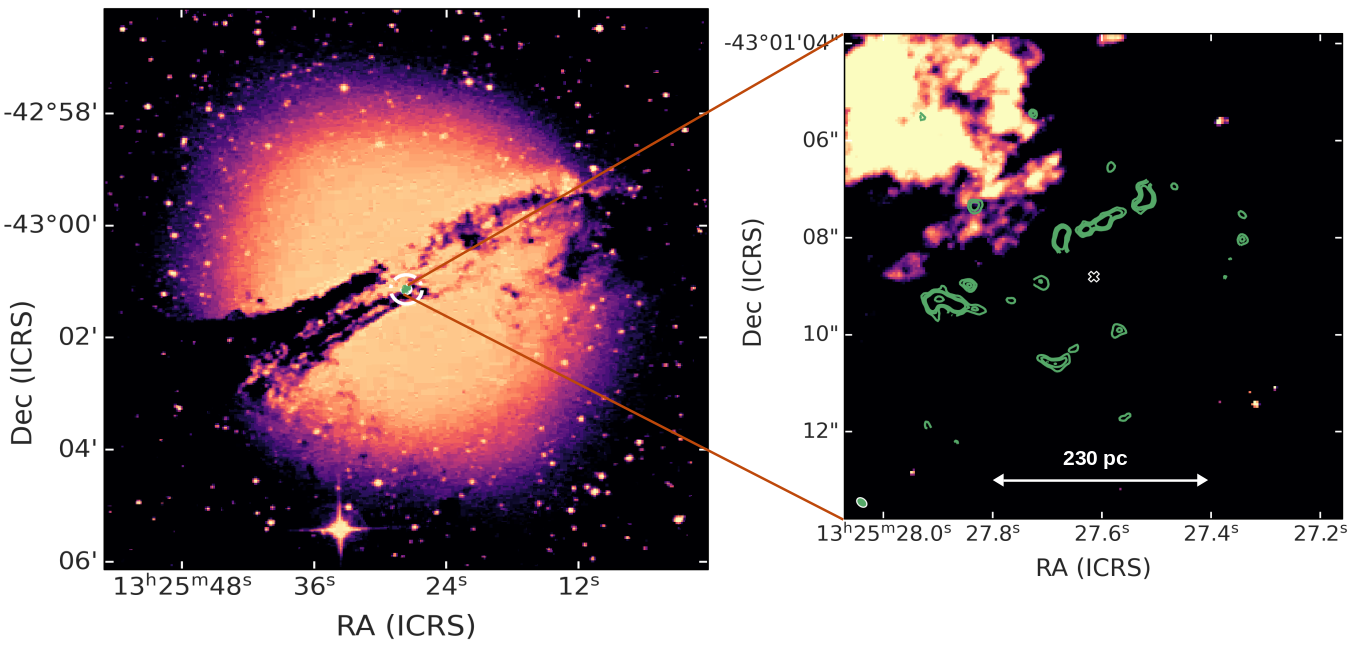}
    \caption{
    \textit{Left panel:} DSS-2 B-band image 
    of NGC~5128.
    The inner white dashed circle indicates the FoV
    of both ALMA observations, with a radius of 4\farcs3 ($\sim$160~pc),
    while the outer dash-dotted circle represents
    the \textit{Herschel}/SPIRE-FTS beam FWHM
    for CO(6--5) observations,
    with a 15\farcs6 radius.
    \textit{Right panel:} zoom of the inner 380~pc,
    with HST WFPC2 F555W image of
    NGC~5128 (from \protect\citealt{marconi00})
    in the background,
    with superimposed, in green,
    the contours of ALMA CO(6--5) moment 0
    at the resolution of 170~mas
    (project 2012.1.00225.S, PI: Espada).
    The contours are drawn at
    (3, 4, 5, 10, 20) $\times \, \sigma$,
    where $\sigma = 0.42$ Jy beam$^{-1}$ km s$^{-1}$.
    The restored ALMA beam
    is shown as a green ellipse with white edges
    at the bottom left.
    A "X" marker, black with white edges,
    indicates the center of the galaxy.
    }
    \label{fig:N5128}
\end{figure*}

\begin{figure*}
    \includegraphics[width=\textwidth]{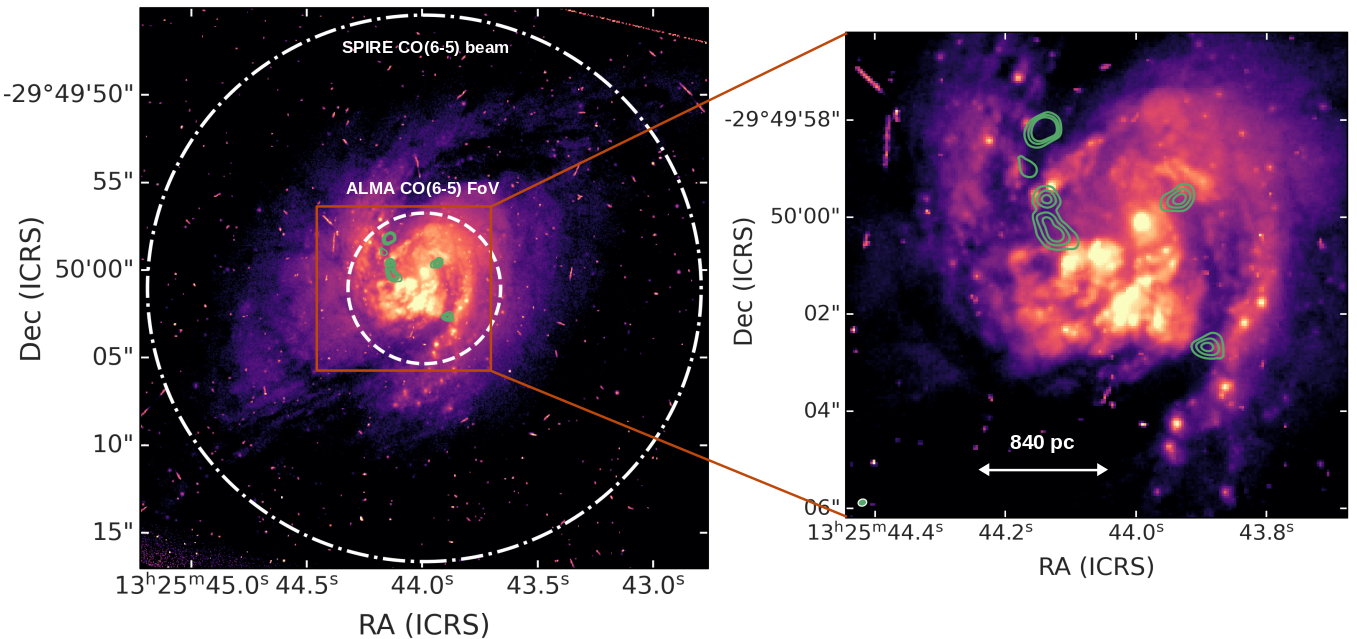}
    \caption{
    \textit{Left panel:} HST WFPC2 F606W image 
    of NGC~5135 (from \protect\citealt{malkan98}),
    with superimposed, in green,
    the contours of ALMA CO(6--5) moment 0
    at the resolution of 170~mas
    (project 2013.1.00524.S, PI: Lu).
    The contours are drawn at
    (3, 4, 5, 10, 20) $\times \, \sigma$,
    where $\sigma = 1.2$ Jy beam$^{-1}$ km s$^{-1}$.
    The inner white dashed circle indicates the FoV
    of both ALMA observations, with a radius of 4\farcs3 ($\sim$1.2~kpc),
    while the outer dash-dotted circle represents
    the \textit{Herschel}/SPIRE-FTS beam FWHM
    for CO(6--5) observations,
    with a 15\farcs6 radius.
    \textit{Right panel:} zoom of the inner 2.5~kpc.
    The restored ALMA beam
    is shown as a green ellipse with white edges
    at the bottom left.
    }
    \label{fig:N5135}
\end{figure*}

\begin{figure*}
    \includegraphics[width=\textwidth]{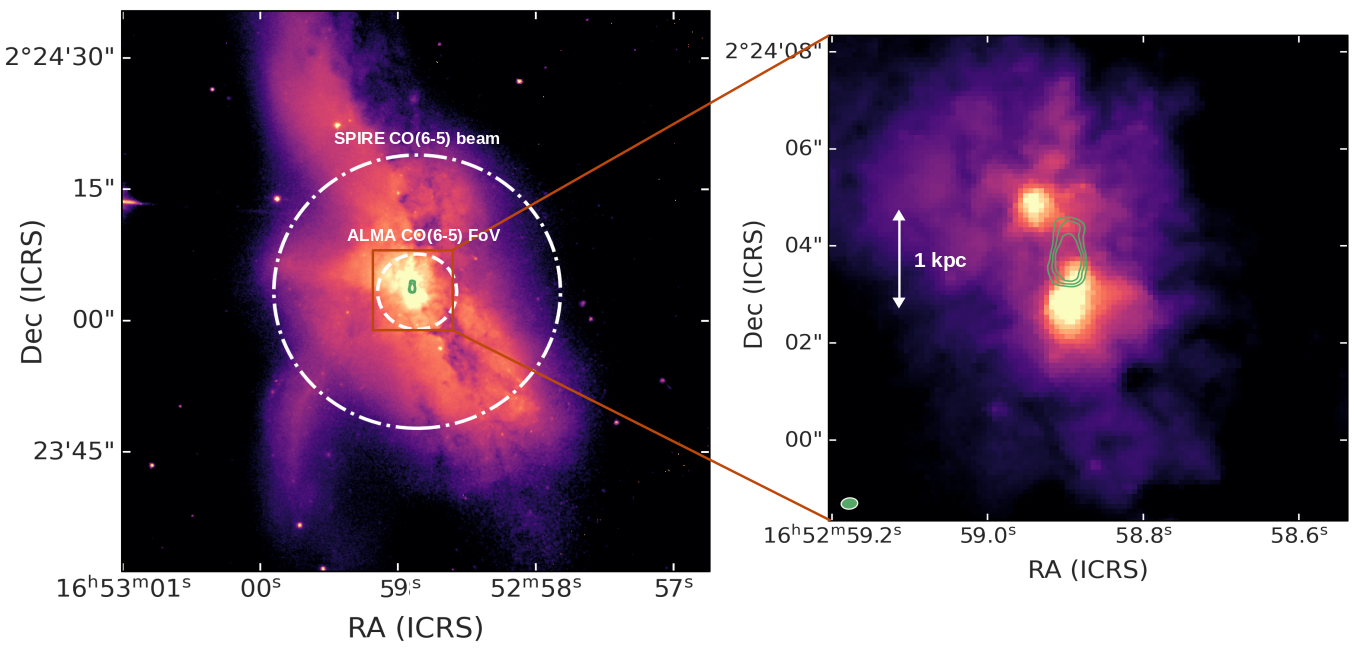}
    \caption{
    \textit{Left panel:} HST WFPC2 F814W image 
    of NGC~6240 (from \protect\citealt{gerssen04}),
    with superimposed, in green,
    the contours of ALMA CO(6--5) moment 0
    at the resolution of 250~mas
    (project 2015.1.00658.S, PI: Rangwala).
    The contours are drawn at
    (3, 4, 5, 10, 20) $\times \, \sigma$,
    where $\sigma = 29$ Jy beam$^{-1}$ km s$^{-1}$.
    The inner white dashed circle indicates the FoV
    of both ALMA observations, with a radius of 4\farcs3 ($\sim$2.1~kpc),
    while the outer dash-dotted circle represents
    the \textit{Herschel}/SPIRE-FTS beam FWHM
    for CO(6--5) observations,
    with a 15\farcs6 radius.
    \textit{Right panel:} zoom of the inner 4.5~kpc.
    The restored ALMA beam
    is shown as a green ellipse with white edges
    at the bottom left.
    }
    \label{fig:N6240}
\end{figure*}

\begin{figure*}
    \includegraphics[width=\textwidth]{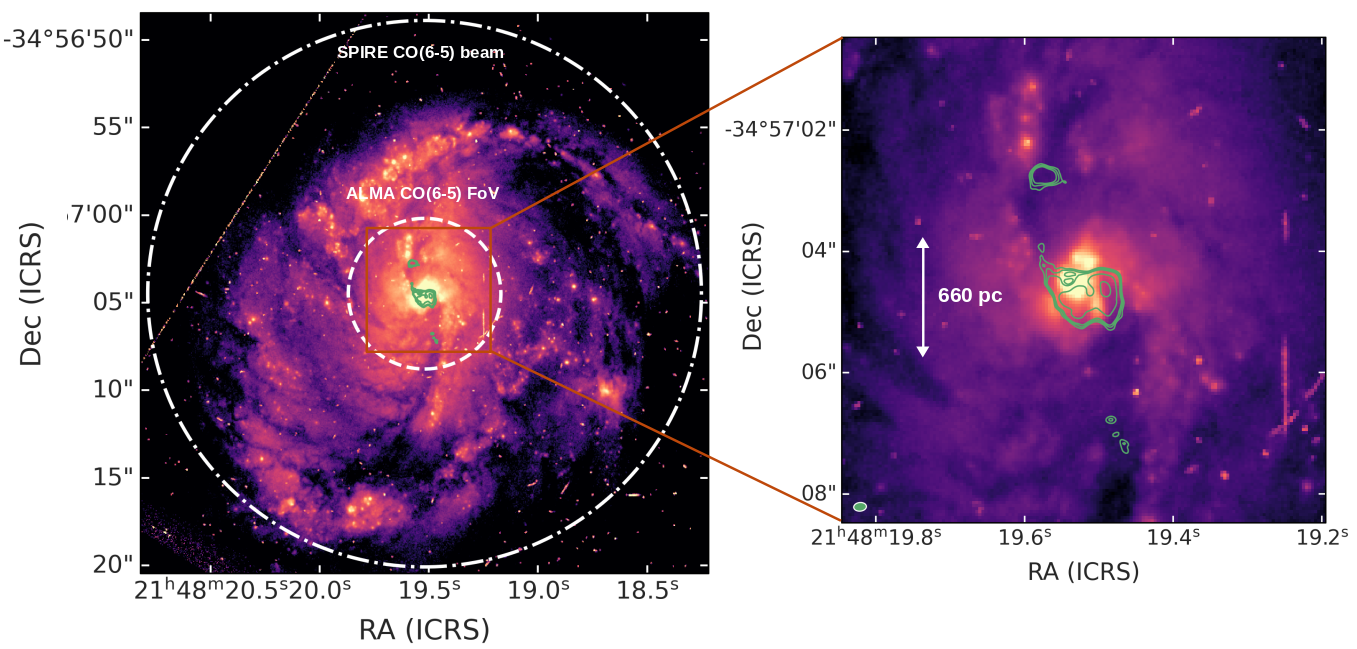}
    \caption{
    \textit{Left panel:} HST WFPC2 F606W image 
    of IRAS~F05189--2524 (from \protect\citealt{malkan98}),
    with superimposed, in green,
    the contours of ALMA CO(6--5) moment 0
    at the resolution of 180~mas
    (project 2013.1.00524.S, PI: Lu).
    The contours are drawn at
    (3, 4, 5, 10, 20) $\times \, \sigma$,
    where $\sigma = 1.5$ Jy beam$^{-1}$ km s$^{-1}$.
    The inner white dashed circle indicates the FoV
    of both ALMA observations, with a radius of 4\farcs3 ($\sim$1.4~kpc),
    while the outer dash-dotted circle represents
    the \textit{Herschel}/SPIRE-FTS beam FWHM
    for CO(6--5) observations,
    with a 15\farcs6 radius.
    \textit{Right panel:} zoom of the inner 2.3~kpc.
    The restored ALMA beam
    is shown as a green ellipse with white edges
    at the bottom left.
    This ALMA image has not been primary-beam corrected.
    }
    \label{fig:N7130}
\end{figure*}


\bsp	
\label{lastpage}
\end{document}